# The Role of the Surface Evapotranspiration in Regional Climate Modelling: Evaluation and Near-term Future Changes

Matilde García-Valdecasas Ojeda[1] (0000-0001-9551-8328),

Juan José Rosa-Cánovas[1] (0000-0001-5320-3109),

Emilio Romero-Jiménez[1] (0000-0002-0572-9103).

Patricio Yeste[1] (0000-0002-0546-9866),

Sonia R. Gámiz-Fortis[1] (0000-0002-6192-056X),

Yolanda Castro-Díez[1] (0000-0002-2134-9119),

and,

María Jesús Esteban-Parra[1] (0000-0003-1350-6150)

[1]Department of Applied Physics. University of Granada

Avda. Campus Fuente Nueva S/N, ES18071. Granada, SPAIN

Correspondence to: Matilde García-Valdecasas Ojeda mgvaldecasas@ugr.es

**Abbreviations**
pr: precipitation
SFCEVP: surface evapotranspiration
SMroot: root-zone soil moisture
T2: near-surface air temperature

# The Role of the Surface Evapotranspiration in Regional Climate Modelling: Evaluation and Near-term Future Changes

Matilde García-Valdecasas Ojeda[1], Juan José Rosa-Cánovas[1], Emilio Romero-Jiménez[1], P. Yeste[1], Sonia R. Gámiz-Fortis[1], Yolanda Castro-Díez[1], and María Jesús Esteban-Parra[1]

[1]Department of Applied Physics. University of Granada, Granada, Spain. mgvaldecasas@ugr.es

**ABSTRACT**

The surface evapotranspiration (SFCEVP) plays an essential role in climate, being the link between the hydrological and energy cycles. Therefore, how it is approximated and its implication in the regional climate are important aspects to understand the effects of climate change, especially over transitional zones such as the Iberian Peninsula (IP). This study aims to investigate the spatiotemporal patterns of the SFCEVP using a regional climate model (RCM), the Weather Research and Forecasting (WRF) model. With this purpose, a set of WRF simulations were completed using different driving data. On the first hand, a recent present (1980-2017) simulation driven by the ERA-Interim reanalysis was carried out to evaluate the suitability of the RCM performance. On the other hand, two global climate models (GCMs) from the CMIP5 initiative, the CCSM4 and the MPI-ESM-LR, were used as driving data to evaluate the GCM-RCM couplings, which is essential to climate change applications. Finally, projected changes were also investigated for a near-term future (2021-2050) paradigm. In general, the results pointed out the WRF model as a valuable tool to study the spatiotemporal patterns of the SFCEVP in the IP, showing an overall and acceptable ability at different spatial and temporal scales. Concerning projections, the results indicate that the IP is likely to undergo significant changes in the SFCEVP in the near future. These changes will be more apparent over the southernmost, and particularly during spring and summer, being in the latter season the SFCEVP fundamentally reduced. These results agree with projected changes in soil moisture, which is probably associated with changes in precipitation patterns. Additionally, the results reveal the major role of SFCEVP in modulating the climate over this region, which is involved in the complex land-atmosphere processes.

***Keywords***: Surface evapotranspiration, land-surface processes, regional climate simulations, Weather Research and Forecasting, Iberian Peninsula.

# 1. Introduction

Surface evapotranspiration (SFCEVP) is a key variable of the state of the soil as it links the energy, carbon, and water cycles (Fisher et al. 2017, Martens et al., 2017). The SFCEVP influences de climate (Dolman et al., 2014) through the occurrence of land-atmosphere feedbacks. These modify precipitation, temperature, humidity, and cloud covers (Seneviratne et al., 2010), and leading to the exacerbation of extreme events such as heatwaves, (Miralles et al., 2014a), floods (Xue et al., 2001), and droughts (Quesada et al., 2012). This fact is particularly true over the so-called transitional zones, where the soil moisture largely controls the climate variability.

Under anthropogenic climate change, the role played by the SFCEVP is even more essential. Increasing greenhouse gas (GHG) concentrations are expected to affect the magnitude of heat fluxes, and their effects will propagate through all the components of the energy and water cycles (Miralles et al., 2016). This fact makes that a better understanding of how this variable behaves under different GHG concentrations be of high relevance for adequately developing mitigation and adaptation strategies for the ongoing climate change. In spite of its recognized importance, the SFCEVP is one of the most uncertain components of the global hydrological balance (Dolman and de Jeu, 2010; Miralles et al., 2016). It is mainly because the current capacity to directly monitor the time course of this variable is unfortunately weak, with limited coverage in time and space of in situ measurements. In recent years, great efforts have been made to develop long-term global evaporative products, such as the Priestley-Taylor model datasets (Fisher et al., 2008), the Global Amsterdam Model datasets (Miralles et al., 2011) or the Global MODIS datasets (Mu et al., 2007). These are the result of applying different algorithms using satellite remote sensing observations such as radiation, precipitation, and soil moisture as input data.

Additionally, climate simulations can be a valuable tool in this context, allowing to achieve long-term variables in a continuous spatiotemporal resolution that could help to understand how the SFCEVP interplays with the atmosphere in both current and future conditions. They provide an overall picture of the soil and atmosphere behaviors through a high number of variables predicted. In this framework, regional climate models (RCMs) were developed to overcome weakness derived from the coarse resolution of the global climate models (GCMs), providing regional climate information at an adequate resolution to study aspects of the climate that require finer-resolution (e.g., land-surface interactions).

In the last decades, RCMs have been widely used to study the spatiotemporal patterns of the current climate

(e.g., Alonso-González et al., 2018; Argüeso et al., 2012a; Politi et al., 2018) as well as to investigate the effects of increased GHGs (Argüeso et al 2012b; Gómez-Navarro et al., 2010; Nguvava et al., 2019). However, only a few studies focused on examining the RCMs performance in terms of variables related with the soil state, and how climate change will influence land-surface processes. In the latter context, Greve et al. (2013) showed the ability of a reanalysis-driven regional simulation to adequately reproduce the root-zone soil moisture over the European region. Knist et al. (2017) found that different RCMs in the framework of the EURO-CORDEX initiative can reproduce the annual cycles of surface fluxes such as the latent and sensible heat fluxes in different European climate zones. For the Spanish region, García-Valdecasas Ojeda et al. (2017) highlighted the capability of regional climate simulations to properly characterize drought spatiotemporal patterns, which are strongly related to land-surface interactions. For the future, Jerez et al. (2012) evidenced the crucial role played by the land-surface models (LSMs) to adequately projecting the climate over the Iberian Peninsula (IP). In another recent study, van der Linden et al. (2019) pointed out the added value provided by an RCM with respect to its driving conditions in projecting soil drying and its potential driving factors in central-western Europe.

This work aims to investigate the ability of a regional climate model, the Weather Research and Forecasting (WRF) model, to characterize the main spatiotemporal patterns of the SFCEVP, an important linking variable between land and atmosphere that has been poorly studied so far. This study was performed over the IP, a topographically complex region characterized by a high spatiotemporal climate variability; thus the use of a regional model is more adequate. To do this, current simulations were firstly generated using WRF in order to evaluate the model performance. How the regional model captures this variable is of high relevance in this region, a mostly transitional zone where land-surface processes largely influence the climate. Moreover, projections of the SFCEVP were also examined for a near future (2021-2050) paradigm using two GCMs from the CMIP5 initiative as forcing data and under two representative concentration pathways (RCPs): a milder scenario (RCP4.5) and the most pessimistic one (RCP8.5). Table 1 shows the global temperature rise projected by the two GCMs used in this study, these being between 1°C and 1.5°C, allowing us to analyze the associated impacts with global warming according to the Paris Agreement (IPCC, 2018). The study was structured as follows: Section 2 describes the data and methodology used in both, the model evaluation and in the assessment of future projections. Section 3 displays the main results achieved, and finally, Section 4 summarizes and discusses the main results of this study.

## 2. Data and Methods

### 2.1. Regional Climate Simulations

The WRF-ARW model (Skamarock et al., 2008) version 3.6.1 was used to generate regional climate simulations over the IP. All runs were completed using the same configuration and they differ only in the data used to force the WRF model.

Firstly, to examine inherent errors associated with the RCM, a simulation driven by the ERA-Interim reanalysis (Dee et al., 2011) was carried out for the period 1979-2017. Additionally, two historical simulations were completed for the period 1979-2005 using as driving data two different GCMs; the bias-corrected CESM1 (Monaghan et al., 2014), and the MPI-ESM-LR (Giorgetta et al., 2013). The latter was previously corrected in systematic bias following the Bruyère et al. (2015) approach, which is the same applied in the CESM1. In this regard, and because the historical simulations end in 2005, both historical simulations were completed until 2017 with the runs driven by anthropogenic climate change under RCP8.5, since it proved to appropriately describe the current climate characteristics (Granier et al., 2011). Additionally, to investigate near-term future changes, regional projections using the above mentioned GCMs were completed from 2020 to 2050 under two RCPs (RCP4.5 and RCP8.5).

Regarding the spatial model configuration, it consisted of two one-way nested domains (Fig. 1): the finer domain (d02) spanning the IP at 0.088° (10 km approximately) of spatial resolution, and nested over a coarser domain (d01) that corresponds to the EURO-CORDEX region (Jacob et al., 2014) at 0.44° (50 km approximately) of spatial resolution. In the vertical, 41 levels were used with the top set to 10 hPa.

One of the most critical steps to adequately configure the WRF model is the selection of the best set of parameterizations for the study region (Argüeso et al., 2011; Jerez et al., 2013; Kotlarski et al., 2014). This is especially important in the case of topographically complex regions such as the IP, so, the parameterizations set was selected according to a previous sensitivity study (García-Valdecasas Ojeda et al., 2015). They are: the Betts-Miller-Janjic (Betts and Miller, 1986; Janjić, 1994) for cumulus, the Convective Asymmetric Model version 2 (Pleim, 2007) for planetary boundary layer, the WRF single-moment-three-class (Hong et al., 2004) for microphysics, and the Community Atmosphere Model 3.0 (Collins et al., 2004) for radiation (long-wave and short-wave). This selected parameterization set has been successfully used to characterize drought patterns over the

Spanish region (García-Valdecasas Ojeda et al., 2017).

Land-surface related variables such as the SFCEVP are achieved by the land surface model (LSM) coupled to WRF. In this study, we used the unified Noah (Chen and Dudhia, 2001) as LSM coupled to WRF (hereinafter referred to as WRF-Noah), which proved to be adequate to simulate the regional climate worldwide. WRF-Noah makes use of different parameters established for the vegetation (e.g., stomatal resistance, leaf area index, etc.) and texture types (e.g., wilting point, field capacity, etc.), which largely control the predicted SFCEVP. In this regard, among the different options provided by WRF, the 21-category MODIS land use from the International Geosphere-Biosphere Programme (IGBP) at a resolution of 30 arc seconds was used, with the soil texture being the default 16-category FAO soil texture.

## 2.2 Reference Data

As reference data, the surface evapotranspiration from the Global Land-surface Evaporation Amsterdam Model (GLEAM) version 3.2a (Martens et al., 2017; Miralles et al., 2011) was used to evaluate the WRF model performance in terms of SFCEVP. GLEAM is a land surface model based on the Priestley and Taylor formulation (Priestley and Taylor, 1972) that provides land evaporation by using remote sensing observations. These data have proved to be noteworthy tools for studying climate variability and trends (Miralles et al., 2014b), but also, more recently, they have been used to evaluate different RCM outputs (González-Rojí et al., 2018; Knist et al., 2017). GLEAM in its version 3.2a is composed by a set of daily data that span the period from 1980 to 2017 in a 0.25º x 0.25º regular grid covering the entire Earth's globe.

Spatiotemporal patterns of the SFCEVP are largely associated with variations in near-surface air temperature (T2) and precipitation (pr), so to gain more confidence in the WRF performance, these two well-known atmospheric variables were also evaluated. To do this, observations from the E-OBS gridded dataset in its ensemble members version 19.0 (Cornes et al., 2018) at 0.1° of spatial resolution was used. E-OBS, created in the framework of the EU-FP6 project ENSEMBLE (Haylock et al., 2008), has proved to adequately represent the main European climate, and now is also available in an improved version resulted from the calculation of an ensemble with 100 members of each daily field.

However, it is worth mentioning that the reference data are also affected by inherent errors, which can be occasionally large. Such errors are unavoidable, so it is essential to consider them as inaccuracies in observations could lead to a misinterpretation in the WRF capability to capture climate behaviors. Concerning the evaporation product used in this study, Miralles et al. (2011) pointed out that GLEAM is highly sensitive to precipitation forcing (Miralles et al. 2011), so errors in the latter are expected to affect the GLEAM performance. In other study, McCabe et al. (2016) found that GLEAM tends to slightly underestimate the evaporation when it is compared with tower-based eddy-covariance observations. Moreover, note that although GLEAM is largely based on observations, it is not strictly observational datasets. Therefore, uncertainties in forcing data must be taken into account together with the sensitivity to parameters associated with the vegetation types, which are different from the WRF-Noah assumptions used in this work for the simulations.

In the same way, uncertainties in observational gridded products can be of similar magnitude as the inherent RCM biases, even in regions where these products are based on dense networks (Gómez-Navarro et al., 2012). Kotlarski et al. (2019) in an exercise of comparison between different gridded and RCM products, found that E-OBS typically underestimate the precipitation and temperature. Likewise, Prein and Gobiet et al. (2017) recognized problems of gridded products such as E-OBS to appropriately capture the amount of precipitation, which can be noteworthy over mountainous areas. This aspect is of high relevance over regions such as the IP, which is characterized by a strong altitudinal gradient (Fig. 1b).

## 2.3. The Model Evaluation

To evaluate the WRF ability to characterize land-surface processes, the SFCEVP, T2, and pr only from the inner domain (d02), and over land were analyzed. The analysis was based on comparing the WRF outputs concerning the reference data from GLEAM and E-OBS for the period 1980-2017. This period was selected in order to perform an evaluation for a climatologically robust period.

Two spatial perspectives were used to evaluate WRF. Firstly, a region-by-region (regional perspective) study was performed. As previously mentioned, the accumulated amount of SFCEVP simulated by WRF depends largely on the vegetation types, so the land-use classification from WRF (Fig. 1S, in supplementary material) was used to select the different regions. In this regard, GLEAM uses a land-cover classification based on four main types (bare soil, short vegetation, tall vegetation, and open water), so with the purpose of performing a more adequate comparison, the land-uses contemplated by WRF were grouped into three main types: tall vegetation (corresponding to evergreen needleleaf forest, evergreen broadleaf forest, and mixed forest), short vegetation

(closed and open shrublands, woody savanna, savanna, grassland, and cropland), and urban region. This classification showed very similar spatial patterns to one achieved using a regionalization procedure (Argüeso et al., 2011) using daily values of SFCEVP from GLEAM (result not shown), suggesting that it is adequate to investigate the model performance from a regional perspective.

The three selected regions were used to obtain the three spatially averaged time series on which the regional perspective was based on. Then, bias, mean absolute error (MAE), and normalized standard deviations (NormStd) were computed to examine the model performance. Also, the model capability to capture the annual cycle of the monthly values of the three variables was explored by regions. Additionally, different simulated percentiles vs. the reference ones through quantile-quantile (Q-Q) plots were represented. The latter analysis allows us to further investigate if WRF can reproduce the probability density functions from the daily reference data. For the daily accumulated pr, the analysis was performed taking into account only those values above 0.1 mm day$^{-1}$, following the methodology proposed by Argüeso et al. (2011).

Secondly, a local perspective (i.e., grid-to-grid comparison) was also used to further explore if WRF reproduces the main spatiotemporal patterns of the SFCEVP, T2, and pr. To make the data spatially comparable, downscaled outputs were remapped onto the GLEAM and E-OBS grids using the nearest neighbor approach. As for the regional perspective, different temporal aggregations were used. Thus, annual and seasonal bias were computed to elucidate the mean deviation for each grid point. The latter time aggregation was also analyzed because authors such as Ruosteenoja et al. (2018) have recently highlighted the importance of studying land-surface processes at seasonal scale as different processes take part along the year. Finally, the WRF ability to reproduce the probability density function of the daily amount of SFCEVP for each grid point was also explored through the Perkins Skill Score (PSS, Perkins et al., 2007).

**2.3. Analysis of the Projections in the SFCEVP**

Changes between the near-term future (2021-2050) and the historical period (1980-2005) for each grid point were examined through their differences expressed in relative terms (percentage). In the same way, changes in the root-zone soil moisture (SMroot; the upper 1 meter of the soil), was also investigated to further analyze the impacts on land-surface processes. To evaluate the significance of these changes, a circular block bootstrap method (Politis and Romano, 1992) are applied using 1000 samples to determine the 95% confidence interval. This method allows taking into account the autocorrelation of the records (Kiktev et al. 2003) as it applies bootstrapping resampling for consecutive records with a given block length (L), instead of individual values. Thereby, significant changes for the future in relation to the historical period can be determined, even for auto-correlated and non-Gaussian data. Here, the circular block resampling was applied following the procedure proposed by Turco and Llasat (2011) that determined L using the method detailed in Politis and White (2004). In this study, L was estimated for each period (annual, DJF, MAM, JJA, and SON) and variable, and the same value of L was used for all grid points. These values, which corresponded to the 90$^{th}$ percentile of all grid points analyzed, ranged from 3 to 10, depending on the period and GCM-driven simulation.

## 3. Results

### 3.1 The Model Evaluation

### 3.1.1. Region-by-Region Analysis

To know how the WRF model captures the main spatiotemporal patterns of the different variables, an analysis of monthly data was firstly performed for every region. Thus, the monthly values for each grid-point were computed, and then, the spatially averaged values for every region was obtained. Table 2 shows the statistic error measurements of the monthly accumulated amount of SFCEVP, monthly-mean T2, and accumulated pr for each region (tall vegetation, short vegetation, and urban). Such measurements were computed for the WRF simulation driven by ERA-Interim (WRFERA), the CESM1 model (WRFCCSM), and the MPI-ESM-LR (WRFCCSM) with respect to the reference data (GLEAM for SFCEVP and E-OBS for T2 and pr, respectively). Note that for the SFCEVP and pr, bias and MAE are expressed in relative terms (simulations *minus* reference data/reference data), meanwhile for T2, these metrics are expressed as absolute differences (simulations *minus* reference data). Both, bias and MAE indicate the averaged deviation in the model concerning the reference data, being the first one also a measure of over- or underestimation. NormStd, however, shows the model behavior in terms of variability. In this regard, positive values indicate that climate variability is overestimated, while negative values show the opposite behavior.

Broadly speaking, WRF captures quite well all variables, except in the case of the SFCEVP over urban regions. In the latter region, all error measurements (bias, MAE, and NormStd) indicate a poor skill concerning GLEAM. For this reason, the results from the SFCEVP over urban regions will be represented hereafter, but these will not be commented. For the other two regions, the SFCEVP shows overestimations, with bias ranging from 0.58 to 17.41. The short vegetation presents lower bias than the tall vegetation, the WRFCCSM being the simulation with the best skill according to this parameter. However, when the WRF simulations are evaluated in terms of MAE and NormStd, the tall vegetation presents a better agreement with the reference data, particularly for the WRFERA simulation. This indicates that the simulations are probably affected by compensation errors, particularly in the case of the WRFCCSM for the short vegetation (bias around 0.6% vs. MAE around 21%).

Unlike for SFCEVP, WRF tends to underestimate the temperature (bias of around -0 .5), except over urban regions. In the latter case, overestimations of around 1ºC appear in all WRF simulations. Additionally, the results indicate that GCM-driven simulations are probably more affected by compensation errors than WRFERA. That is, while the WRFERA presents values of similar magnitude for bias and MAE, the WRFCCSM and WRFMPI show higher differences between these two metrics. In terms of variability, however, all WRF simulations present a good skill, especially for the short vegetation, and greater for the WRERA and WRFCCSM simulations (NormStd close to 1).

The results also show that the precipitation is typically overestimated. This behavior is more apparent for the tall vegetation, and especially for the WRFMPI simulation (wet-bias of around 50%). Moreover, the higher the precipitation errors, the greater the deviations from the SFCEVP. Therefore, this evidences the relationship between the model performances in terms of these two variables. As shown in the NormStd, the precipitation variability is mostly overestimated by WRF, particularly over the tall vegetation. In this regard, note the number of grid-points representing each region, which is much less for the tall vegetation.

Fig. 2 shows the annual cycle of the monthly amount of SFCEVP, T2, and pr from reference data and for all WRF simulations (WRFERA, WRFCCSM, and WRFMPI). In general, WRF presents a good skill to capture the overall shape of the annual cycle of all variables analyzed in this study. The largest differences concerning the reference data are shown for the tall vegetation. For this region, and in terms of SFCEVP, the WRFERA presents a generalized overestimation, especially in winter (December-February) and the late summer (i.e., June-July). Concerning results from GCM-driven simulations, while the WRFCCSM behaves similarly to WRFERA (showing even a better agreement with GLEAM for June), the WRFMPI shows larger differences with respect to GLEAM, particularly during the second part of the year.

For T2, however, all WRF simulations are very similar, being this variable overall underestimated, especially during summer. The behavior is probably associated with a large amount of precipitation simulated by WRF during the preceding months (see annual cycles for pr in spring). It leads to an overestimation in the soil water available to evapotranspiration, and subsequently the underestimation in the T2. Contrariwise, the T2 is systematically overestimated for the urban region, as indicated by Table 2. In terms of precipitation, the results present more discrepancies with the reference data. For the tall vegetation, the greatest precipitation deviations appear from October to May, when the highest precipitation occurs, these being again more apparent for the WRFMPI. Similar conclusions can be drawn for the short vegetation. In this region, underestimations for May-June and September-October are shown, for the WRFERA and especially for the WRFCCSM. For the rest of the year, and for the WRFMPI, however, the SFCEVP is slightly overestimated, which again coincides with a generalized overestimation in the pr, and underestimation in the T2.

Fig. 3 displays the simulated percentiles ($25^{th}$, $50^{th}$, $75^{th}$, $80^{th}$, $85^{th}$, $90^{th}$, $95^{th}$, and $99^{th}$) of the daily SFCEVP, T2, and pr *vs.* the observed ones through a Q-Q representation. Gray line indicates a perfect agreement with the reference data, providing a division between overestimated and underestimated percentiles. In general, WRF is able to capture the daily probability distribution of the reference data for all variables. For the tall vegetation, the SFCEVP distribution is slightly overestimated, especially for the larger daily evapotranspiration rates, and higher for the WRFMPI. However, for the short vegetation, the overall agreement with GLEAM is really good, showing the WRFERA and WRFCCSM slight underestimations in the upper percentiles. However, the WRFMPI slightly overestimates all the percentiles except to the $95^{th}$ and the $99^{th}$. Consistently, similar results between regions and simulations are shown in terms of T2, which is, in general, slightly underestimated, except for the urban regions. In terms of precipitation, and for the tall vegetation, the simulations present similar distributions, with the light precipitations being underestimated with respect to E-OBS. The WRFMPI, however, tends to show a higher precipitation amount than the reference data, especially in the upper-percentiles, showing thus, overestimations. For the short vegetation and urban regions, the precipitation is usually underestimated, except for the WRFMPI in the $99^{th}$ percentile for the short vegetation.

**3.1.2. Grid-by-grid Analysis**

Fig. 4 displays the WRF SFCEVP deviations with respect to GLEAM for a grid-point perspective. Annual

(January-December), winter (December-February, DJF), spring (March-May, MAM), summer (June-August, JJA), and fall (September-November, SON) biases are displayed for the three WRF simulations (WRFERA, WRFCCSM, and WRFMPI), which are expressed in relative terms (%). Additionally, to determine the spatial agreement between the averaged patterns of the SFCEVP, pattern correlations (r), which are the spatial correlation between the observed and simulated mean values, are displayed in the bottom right corner of each panel. Due to the WRF anomalous behavior on urban grid-points, they are not represented in this analysis.

In general, WRF represents the spatial patterns of the annual amount of SFCEVP with admissible accuracy in most of the IP, showing pattern correlations up to 0.75 (Fig. 4). WRFERA broadly captures the main GLEAM climatological features, locating the highest SFCEVP (around 800 mm/year in both GLEAM and WRF) over the northernmost IP, and the lowest ones (below 250 mm/year) in the southeastern IP. However, slight overestimations are observed in some parts of the Northern Plateau, where positive deviations up to 75% are reached. Additionally, the model underestimates the annual SFCEVP over northern Portugal, showing negative differences up to 50% in all WRF simulations. Concerning differences between the simulations, and as shown in the regional perspective, the WRFCCSM achieves similar features to WRFERA, while WRFMPI overestimates the annual SFCEVP (large regions present biases up to 75%-100%).

However, the model behaves differently throughout the year. Thus, during winter, when GLEAM shows the lowest amount of SFCEVP, (showing values below 100 mm, Fig. 2S), the WRF overestimations are generalized, reaching biases around 75% in a large part of the IP. In this regard, it is important to keep in mind that errors here are expressed in relative terms, so admissible values in absolute terms may lead to large differences in relative terms. Additionally, all WRF simulations show similar spatial patterns of bias, this being greater in magnitude for the WRFMPI. The largest overestimations appear over the Ebro River Basin, Balearic Islands, and some coastal regions (e.g., the Cantabrian coast), where differences with respect to GLEAM above 150% are reached. The latter behavior probably results from differences between the resolutions, being thus, the definition of the coastal borders different between the different data sets.

During summer, GLEAM presents the most marked northwest-southeast gradient with SFCEVP ranging from 20 to 500 mm (Fig. 2S, first column, JJA). Thus, the highest evapotranspiration rate appears over the northernmost of the IP, where the soil water available is not limited, and thus, the temperature rise results in more

SFCEVP. Contrariwise, the rest of the IP presents a soil moisture-limited regime, meaning that the soil water available to evaporate is scarce during this season, and then, the SFCEVP is mainly constraint. In general, WRF reproduces quite well this feature, with patterns correlations being above 0.75 in all WRF simulations (Fig. 4, JJA). However, certain discrepancies appear in regions where GLEAM indicates really low SFCEVP values (Fig. 2S, first column, JJA, southeastern IP), showing biases above 175%. Also, all WRF simulations present underestimations (e.g., western IP and Balearic Islands), reaching negative deviations of around 75-100%.

The best agreement between GLEAM and WRF occurs in the intermediate seasons (Fig. 4, MAM and SON), showing differences with respect to GLEAM below 25% in most of the IP. Thus, spring SFCEVP is relatively high (Fig. 2S, first column, MAM), result from the increase in temperature in a season when the soil water is still enough. In this framework, WRF seems to show a remarkable ability to capture these features, especially for the reanalysis-driven simulation. For this season, the highest differences regarding GLEAM are presented in the Northern Plateau, these being of around 25-50% in all WRF simulations. By contrasts, underestimations occur over the Pyrenees, Balearic Islands, Portugal, and across some coastal regions. Again, very similar results to WRFERA are found for the WRFCCSM, presenting the WRFMPI a higher presence of overestimations (up to 50%), particularly over the Guadalquivir Basin. For fall, the evapotranspiration is low in practically all the IP (SFCEVP below 200 mm, Fig. 2S, first column, SON), being this behavior the best represented in the simulation driven by ERA, where pattern correlation of 0.78 is shown (Fig. 4). However, it is worth mentioning that WRF also presents some difficulties, showing both overestimations (e.g., the Cantabrian Coast) and underestimations (e.g., the northern Portugal and southern IP). For the WRFCCSM, broader areas than WRFERA present underestimations, showing most of Portugal deviations up to -75%. The WRFMPI, however, as for the other seasons, presents a generalized overestimation pattern (bias about 75-100%) in those areas where the other simulations broadly capture the SFCEVP from GLEAM.

Annual and seasonal WRF T2 deviations with respect to E-OBS are shown in Fig. 5. All WRF simulations present a remarkable ability to represent the spatial patterns of T2 throughout the year (Fig. 2S, second column), showing all simulations pattern correlations of 0.97 (or higher) in all seasons (Fig. 5). At an annual scale, very similar results are found for all WRF simulations, showing a generalized cold-bias of around 1-1.5ºC. The underestimations are particularly marked at high altitudes, where differences with respect to E-OBS up to -2.5ºC are reached over the Pyrenees. The cold-bias presented at annual scale remains throughoutthe year. In this regard, the highest deviations (cold-biases below -2.5ºC) appear over the Pyrenees during winter and spring, and along the Portuguese coasts in summer, the latter particularly shown in the WRFERA and WRFCCSM. The results also reveal that the WRFMPI presents a generalized underestimation in T2 during spring, fall, and especially in summer. In this season, biases below -1ºC occurs in practically all the IP. However, the WRFMPI presents a highlighted agreement with E-OBS during winter, being even better than those from the WRFERA and WRFCCSM. Certain overestimation is also found in the simulations, more apparent during summer when warm-bias up to 2.5ºC appear in the northeastern and the southernmost IP. Also, a generalized overestimation occur over those grid-points that represent urban regions in agreement with the results from the regional perspective.

Analogously, Fig. 6 displays the precipitation bias expressed in relative terms (%) at annual and seasonal time scales. Despite the broad WRF performance in terms of precipitation is quite good (pattern correlations above 0.7), all WRF simulations consistently show overestimations with respect to E-OBS. These are especially highlighted at high altitude, and overall during winter. The spatial patterns of the precipitation bias present some similarities with those from the SFCEVP (Fig. 4), suggesting that inaccuracies in SFCEVP could be partly associated with errors in precipitation. For instance, overestimations are found over the Northern Plateau in practically all the periods analyzed (i.e., annual, DJF, MAM, JJA, and SON) and in all simulations, being this pattern also presented in the SFCEVP (Fig. 4). Similar conclusions can be drawn through the results in the fall biases and by the marked overestimations appeared in summer in the Sierra Nevada (Baetic System), in the south of the IP. Here, the highest summer overestimations appear, in both SFCEVP and pr. Moreover, both variables show the largest differences with respect to the reference data during winter as for SFCEVP.

Finally, the ability of WRF simulations to represent the daily distribution of the SFCEVP was also examined using the PSS (Fig. 7). This was computed by grouping the daily SFCEVP using 19 bins according to the range of values of each grid-point from the GLEAM datasets. PSSs of 100% indicates a perfect fit between the WRF simulations and reference data, meaning a value of 0% that the modeled and reference data are totally different in their daily distributions. The PSS reaches the maximum values (above 90%) over the Guadalquivir and Guadiana River Basins, particularly for the WRFERA and WRFCCSM simulations. As already mentioned, coastal regions in all simulations show important discrepancies between WRF and GLEAM, with PSS values of around 10%. Also, low PSS values appear over the eastern part of the IP in all simulations, reaching values of around 65%. However, in general terms, it can be seen as WRF simulations present a satisfactory agreement with GLEAM in terms of SFCEVP daily distribution.

**3.2. Near-term Changes in SFCEVP**

Once the WRF capability to adequately characterize the main spatiotemporal patterns of the IP has been evidenced, this section is devoted to analyzing the near-term future predictions in the SFCEVP. Fig. 8 shows annual and seasonal SFCEVP changes projected for the period 2021-2050 with respect to the corresponding historical conditions (1980-2005), expressed in relative terms. In columns, the WRFCCSM (first and second columns) and the WRFMPI (third and fourth columns) simulations under the two RCPs (RCP4.5 and RCP8.5) were represented. Black dots indicate non-significant changes at the 95% confidence level. Also, the spatially averaged change for the whole IP is indicated in the bottom right corner of each panel.

Most of the IP is likely to undergo reductions in the annual SFCEVP, which could be, on average, of around 2% for the WRFCCSM, and about 5% and 8% for the WRFMPI under RCP4.5 and RCP8.5, respectively. The highest diminutions are projected by the WRFMPI simulations, where significant differences concerning the historical values are shown in large part of the IP. All WRF simulations consistently indicate that the most affected region will be the southern IP, where the SFCEVP could be reduced up to 15%. Additionally increases in evapotranspiration are also shown over high-altitude regions such as the Cantabrian Ranges and the Pyrenees, where the SFCEVP is projected to increase up to 5% and 15%, respectively. When these results are compared with the projections in precipitation (Fig. 3S in supplementary material), it can be seen the variations in SFCEVP are probably influenced by changes in pr, showing both very similar spatial patterns of changes. Additionally, a common spatial behavior of the SFCEVP changes with those from the T2 (Fig. 4S in supplementary material) is shown, also suggesting the relationship between the changes in both variables. That is, the greater the reductions in SFCEVP are, the stronger the temperature rise in general terms. The latter suggests that the IP could experience a major control of the soil moisture conditions via land-atmosphere feedbacks. An opposite behavior, however, is shown over regions such as the Pyrenees, where increases in both variables are projected.

The evapotranspiration over the IP presents marked differences throughout the year (Fig. 2S), so different implications of the rising GHG concentrations are expected at a seasonal time scale. During winter (Fig. 8, DJF), significant positive deviations with respect to the historical conditions appear in different regions, with the maximum increases being in the southeastern coasts, and over high-altitude regions in the northernmost (e.g., the Cantabrian Range). Here, SFCEVP increases above 15% are reached in all simulations except for the WRFCCSM RCP4.5. Such increases occur together with an enhancement of the precipitation (Fig. 3S) except for the Pyrenees. Increases in SFCEVP over the Pyrenees appear stronger during spring (Fig. 8, MAM) when differences with respect to the historical period above 45% appear under RCP8.5. The latter coincides with a marked warming rate (Fig. 4S) together with non-significant changes in precipitation (Fig. 3S). Therefore, in this case, the temperature rise seems to be a driving factor of changes in evapotranspiration. Also, for this season, and over the northernmost IP, the WRFCCSM projects evapotranspiration increases (around 5%), while the WRFMPI shows some regions with a reduction of this variable under RCP4.5, which are greatly extended under RCP8.5. By contrast, reductions up to 15% are presented over the southernmost IP for all MAM projections.

The most dramatic reductions of SFCEVP are projected in summer (Fig. 8, JJA). For this season, the spatially averaged changes are around -9% for both WRFCCSM simulations and the WRFMPI RCP4.5, reaching -12% for the WRFMPI RCP8.5. Again, the southernmost IP is the most affected, where decreases regarding the historical period are up to 40% over the Guadalquivir Basin. By contrast, all WRF the simulations show SFCEVP increases up to 10% over the Pyrenees. For fall, the results are more uncertain, showing the simulations more differences in their patterns of change. That is, while the WRFCCSM indicates significant increases, especially over the Ebro River Valley, Balearic Islands and across the southeastern coasts, the WRFMPI reveals a generalized reduced SFCEVP, more apparent under RCP8.5.

To further investigate the SFCEVP changes behavior, changes in soil moisture have been also analyzed. Fig 9 shows the projections in the SMroot for the period 2021-2050 with respect to the historical one (1980-2005), expressed in relative terms (%). At annual scale, and consistently with the changes in the SFCEVP, the SMroot is likely to suffer significant decreases showing both similar spatial patterns of changes (spatially averaged diminutions between 2% and 3% for the WRFCCSM and about 3.5% and 7% for the WRFMPI under RCP4.5 and RCP8.5, respectively). During winter, SMroot increases appear in the southeastern coasts, showing increments up to 20% (Fig. 9, DJF). Also, the pr (Fig. 4S, DJF) is projected to increase over the same region, so the results are suggesting the latter as the cause of the increase in SFCEVP. By contrast, during spring, part of the regions where the SFCEVP is increased (i.e., the northernmost IP, and especially the Pyrenees) shows a diminution in the SMroot (non-significant in many cases), indicating thus the SFCEVP as a potential soil-drying driver. For summer (Fig. 9, JJA), reductions in SMroot are mostly generalized (values of around -15%). As for SFCEVP, more discrepancies are shown during fall, although a general soil trend appears with reductions of around 5%.

## 4. Discussion and concluding remarks

This work aims to investigate the WRF model performance in terms of surface evapotranspiration, an essential variable that has been poorly studied, mostly due to the lack of long-term data regular in space and time, and therefore, how the WRF model behaves in this sense remain uncertain.

Consistent with previous studies (Knist et al., 2017; Greve et al., 2013), the WRF model presents a good ability to represent land-surface processes, thus being, a valuable tool to achieve climate information to investigate spatiotemporal patterns of the SFCEVP. Exceptions are the urban grid-points, where WRF showed a poor skill to represent the SFCEVP. This feature agrees with previous studies (González-Rojí et al., 2018; Knist et al., 2017), and is probably related to an anomalous WRF behavior associated with the mismatch between the real land use and the simulated one. Therefore, and with the exception mentioned, the amount of SFCEVP has been satisfactory represented at all the time and spatial scales analyzed (from annual to daily time scales and from regional to local scale). This is especially good for intermediate seasons (i.e., spring and fall) when important biological processes occur, and therefore, its adequate representation is crucial.

However, some discrepancies with respect to GLEAM appear in our simulations, particularly for the WRFMPI simulation. In this regard, it is important to keep in mind that GLEAM is a model based on satellite forcing data and not a direct result from observations. Therefore, part of the differences here found may be due to differences in the vegetation types used by WRF and GLEAM, the different spatial resolutions, how both models represent the soil water availability, and the different parameters associated to the vegetation types (e.g., root depth), and soil texture (e.g., field capacity and wilting point).

The model performance to correctly represent these variables is largely influenced by errors in other atmospheric variables and vice versa. In this regard, it is well-known that the SFCEVP is mostly influenced by precipitation and radiation (and therefore temperature). In this way, the results suggest that a part of the problems to simulate the amount of SFCEVP is associated with the model ability to capture precipitation patterns. That is,

WRF overestimates the precipitation where the simulated SFCEVP is also higher than in GLEAM, leading to greater soil water availability, and thus, more evapotranspiration. In this regard, the largest differences in terms of precipitation with respect to the reference data appeared during winter when the precipitations are largely controlled by the large-scale circulation patterns. The latter agree with the results found by Argüeso et al. (2012a), who indicated that part of the errors in the precipitation simulated by WRF are inherited from the driving data during this season. Additionally, it should be noted that the reference data are not error-free, so uncertainties in both, SFCEVP and pr, could be actually smaller due to the fact that the products used in this study to validate WRF are not fully observational. For instance, large overestimations in precipitation occur at high altitude. In this regions, the gridded product are typically affected by underestimations mainly because observational stations are scarce and the spatial heterogeneity is higher. The precipitation patterns here shown agree with other studies performed over the IP. For instance, Argüeso et al. (2012a) and Herrera et al. (2010) reported higher spreads in spring rainfall by simulating the climate over the Spanish territory climate using regional climate simulations. On the other hand, our findings for a regional perspective agree with those found by Jiménez-Guerrero et al. (2013), who found underestimations over the southernmost IP and along the Mediterranean coast, especially during fall using RCMs simulations driven by ERA-Interim.

Also, the results indicate a generalized underestimation of the temperature, which agrees with other studies performed in the framework of the ESCENA and EURO-CORDEX initiative for our study region (Katragkou et al., 2015). Such a behavior is not just a characteristic of WRF but also of others RCMs (Jiménez-Guerrero et al., 2013; Kotlarski et al., 2014), which in part could be attributable to the overestimated soil water available in this region (result not shown). Thus, under higher than "real" water availability, more latent heat fluxes, and then, less sensible heat fluxes occur, with the subsequent overestimation in temperature. The latter is corroborated by the results obtained over the urban grid-points, where T2 is overestimated at the different time scales analyzed. Therefore, this study could be evidencing the essential role of the SFCEVP on changes in the variability of T2. Actually, anomalous latent heat fluxes favor the enhancement of the sensible heat fluxes, which in turn, lead to more temperature.

In the context of a global increase in the temperature of around 1 and 1.5ºC, changes in SFCEVP with respect to the historical period are shown from all simulations throughout the year. The results also show that model uncertainties are higher than those from different scenarios, as evidenced by Hawkins and Sutton (2009) in their study of the potential uncertainties in climate predictions. In this regard, although differences between GCM-driven simulations occur, common change trends in the SFCEVP appear for all WRF simulations. Thus, the IP is likely to undergo significant reductions in SFCEVP, generalized for nearly all the IP during summer, and over the southernmost in spring. This behavior could be the result of the ongoing soil drying, which seems to be mostly caused by changes in precipitation patterns. Furthermore, the results seem to indicate certain amplification in the temperature rise via positive temperature-soil moisture feedbacks. Over the northernmost, however, enhanced SFCEVPs during spring could compensate for the temperature rise (cooling effect), being thus a soil drying driver as shows the SMroot projections in this region. Interestingly, a common noteworthy increase of the SFCEVP is found over the Pyrenees, particularly apparent during spring and summer. Here, the soil water availability is likely to increase leading to more SFCEVP. This feature, probably caused, at least in part, by the snow-cover depletion (Rangwala and Miller, 2012), could further alter the interactions between land and atmosphere (Xu and Dirmeyer, 2012). All these results evidence the major role of the changes in SFCEVP, which could alter the entire climate system over the IP, a transitional region with a climate largely controlled by the land-surface interactions. These changes could lead to important implications on several natural and social systems through alterations of the hydrological cycle.

Simulations of the U.S. Hydrological Cycle: A Case Study of the 1993 Flood Using the SSiB Land Surface Model in the NCEP Eta Regional Model. Mon. Weather Rev. 129, 2833–2860. https://doi.org/10.1175/1520-0493(2001)129<2833:TIOLSP>2.0.CO;2

## Acknowledgments

This study was financed by the Spanish Ministry of Economy, Industry and Competition, with additional support from the European Community Funds (FEDER) [CGL2017-89836-R]. We thank the anonymous reviewers for their valuable comments that helped to improve this study. We thank the ALHAMBRA supercomputer infrastructure (https://alhambra.ugr.es) for providing us with computer resources.

**Figure Captions**

**Fig. 1** (a) Mean topographical features in the IP and (b) the studied region corresponding to a two nested domain: d01- the EURO-CORDEX region at 0.44º of spatial resolution and the d02 centered over the IP at 0.088º of spatial resolution.

**Fig. 2** Annual cycle of monthly amount of accumulated SFCEVP (first row), average T2 (second row), and accumulated precipitation (third row) for the different WRF simulations and the reference data for the period 1980-2017 in the three study regions (tall and short vegetation, and urban region).

**Fig. 3** Percentiles ($25^{th}$, $50^{th}$, $75^{th}$, $80^{th}$, $85^{th}$, $90^{th}$, $95^{th}$, and $99^{th}$) simulated by the different WRF simulations (WRFERA, WRFCCSM, and WRFMPI) of the daily distributions of the SFCEVP (first row), T2 (second row), and pr (third row) *vs.* those from reference data (GLEAM for SFCEVP and E-OBS for T2 and pr) for the period 1980-2017. The columns comprise the different study regions (tall and short vegetation, and urban region). Gray line indicates a perfect agreement with the reference data.

**Fig. 4** Annual and seasonal relative bias of the amount of SFCEVP for the WRF simulations (WRFERA, WRFCCSM and WRFMPI) with respect to the reference data (GLEAM). Pattern correlation are indicated in the bottom right corner of each panel.

**Fig. 5** Annual and seasonal bias of T2 for the WRF simulations (WRFERA, WRFCCSM and WRFMPI) with respect to the observations from E-OBS. Pattern correlation are displayed in the bottom right corner of each panel.

**Fig. 6** As Fig. 4 but for the accumulated precipitation (pr). Bias is expressed in relative terms with respect to the observations.

**Fig. 7** Perkins Skill Score (PSS) expressed in percentage for the simulated (WRFERA, WRFCCSM, and WRFMPI) daily distribution of the amount of SFCEVP with respect to the reference data (GLEAM).

**Fig. 8** Near future-to-present changes of the amount of SFCEVP expressed as relative differences (future *minus* present/present) for the WRFCCSM and the WRFMPI simulations and under the two RCPs (RCP4.5 and RCP8.5). Non-significant changes at the 95% confidence level are marked with black dots. The spatial averaged change for the whole IP is indicated in the bottom right corner of each panel.

**Fig. 9** As Fig. 8 but for the root-zone soil moisture.

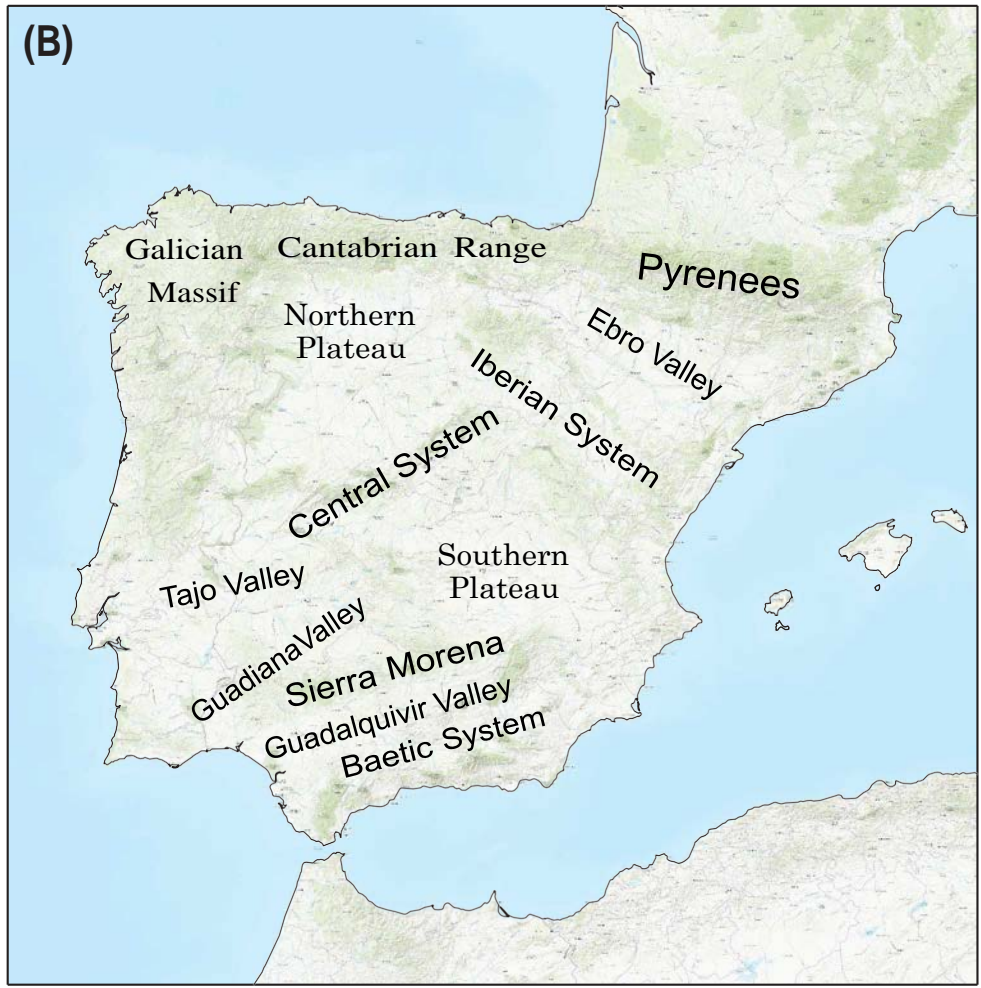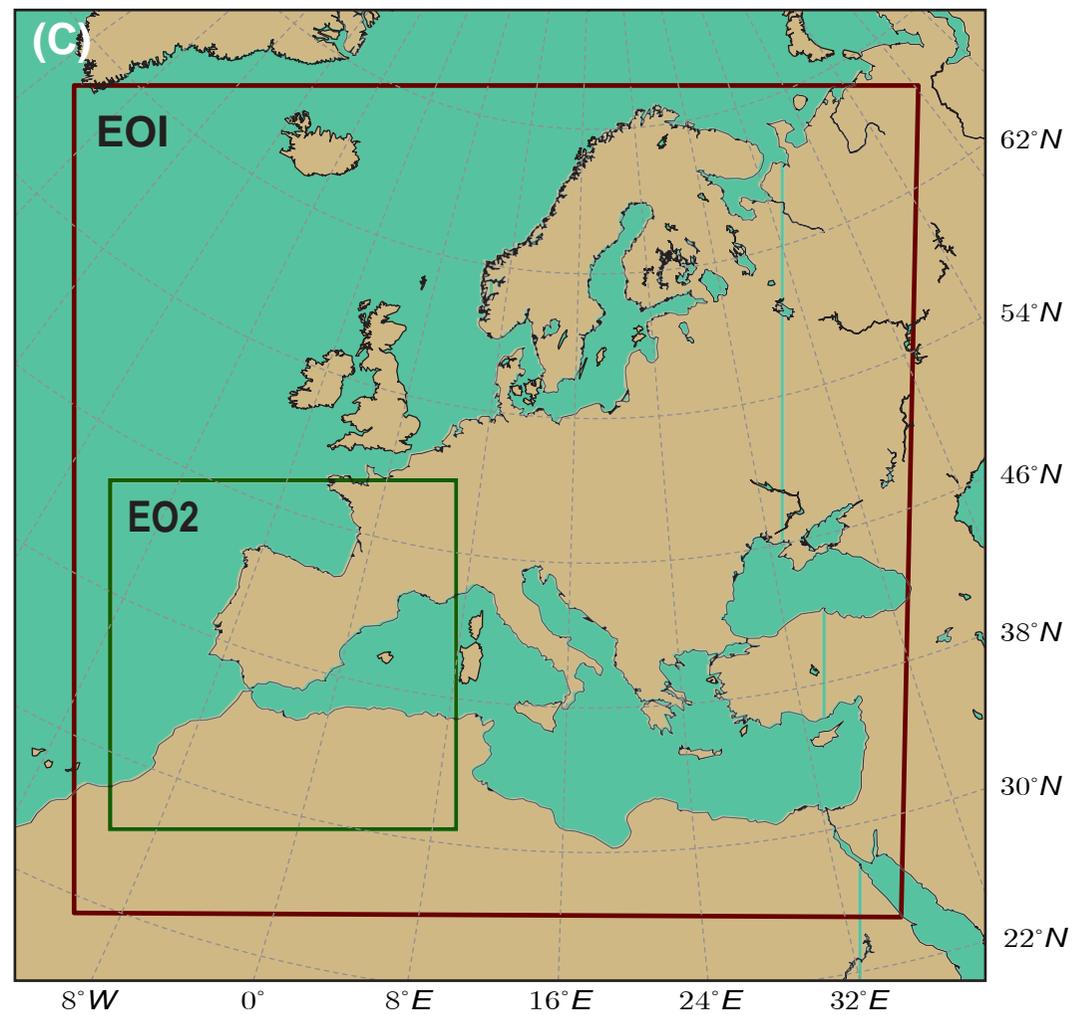

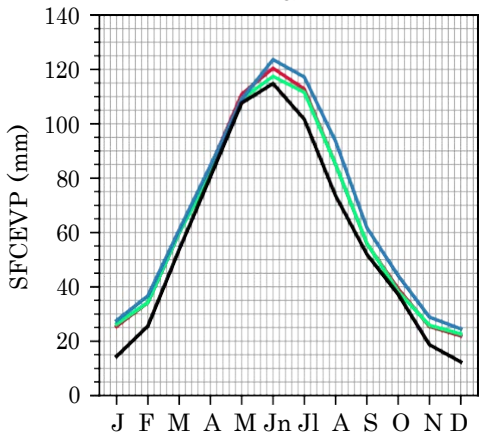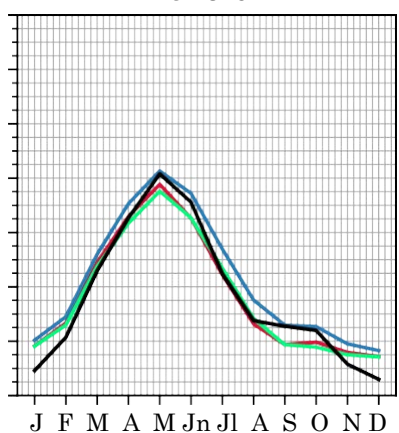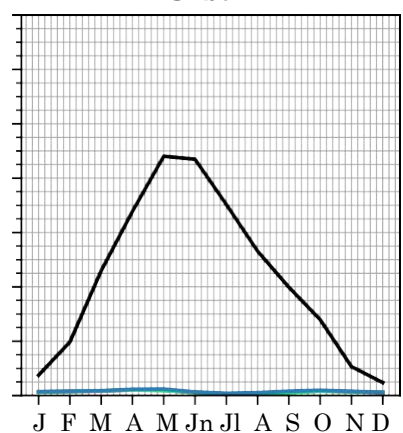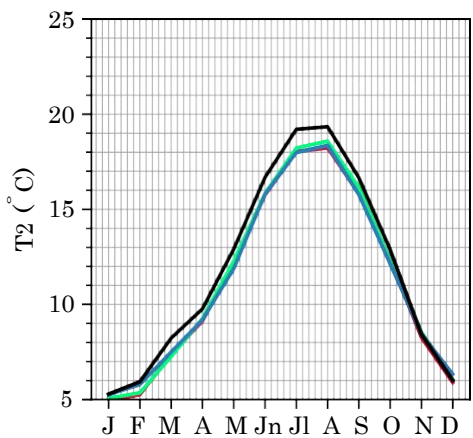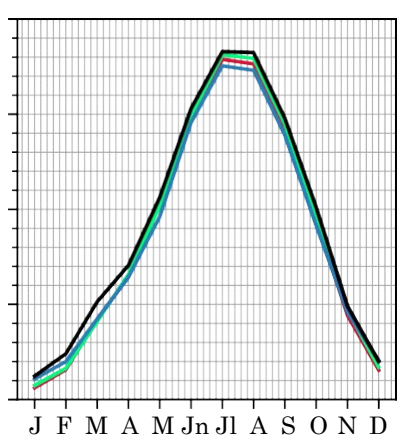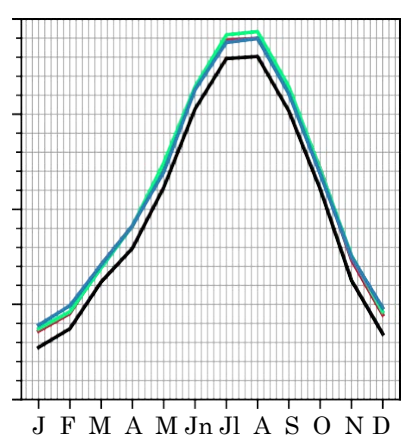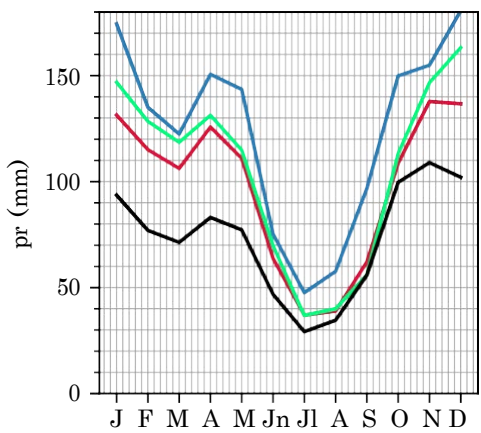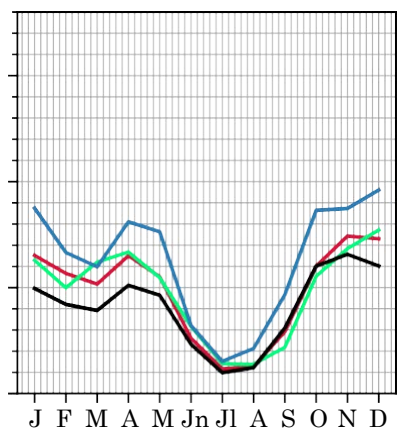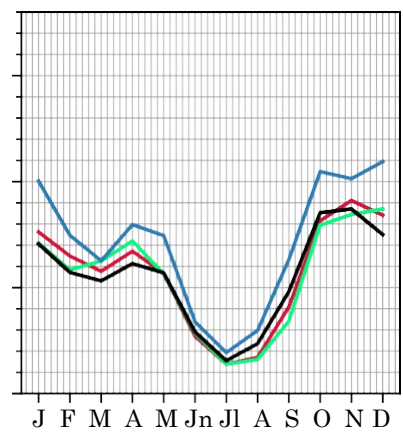

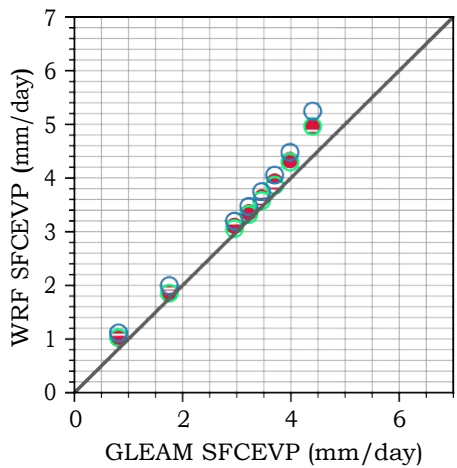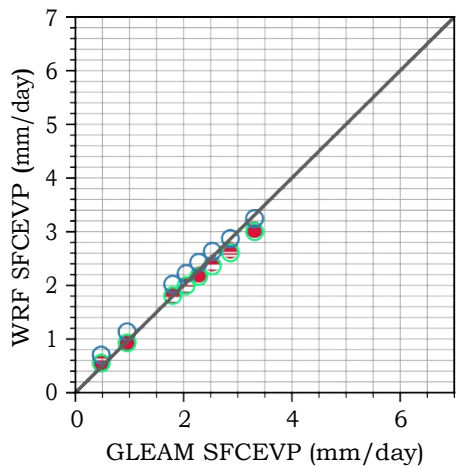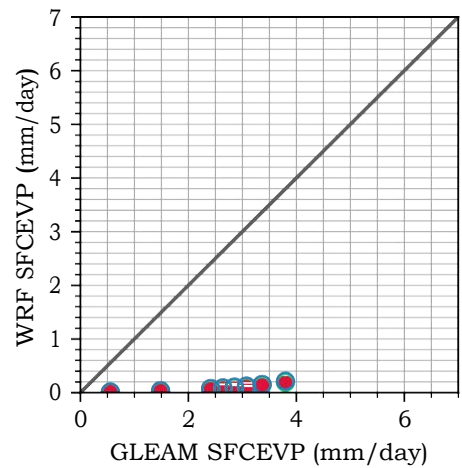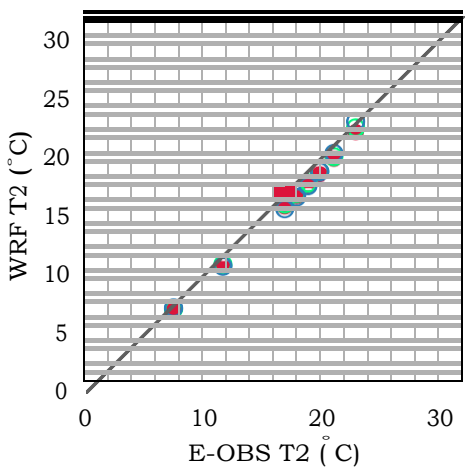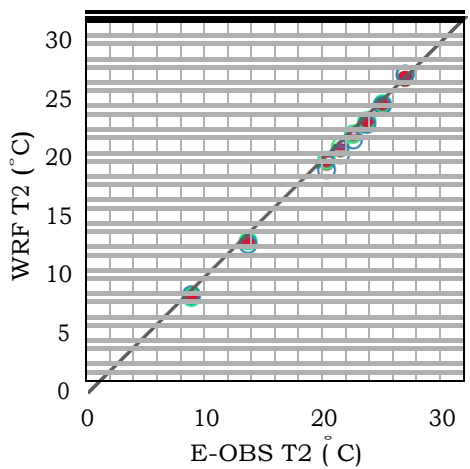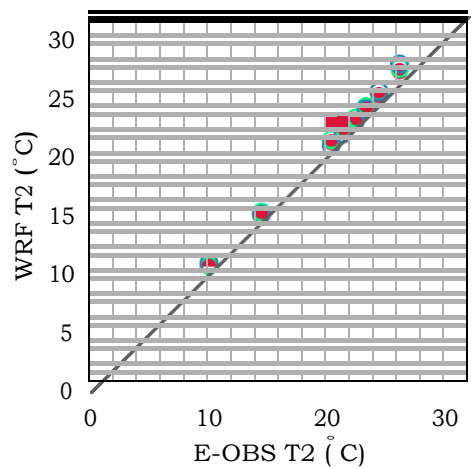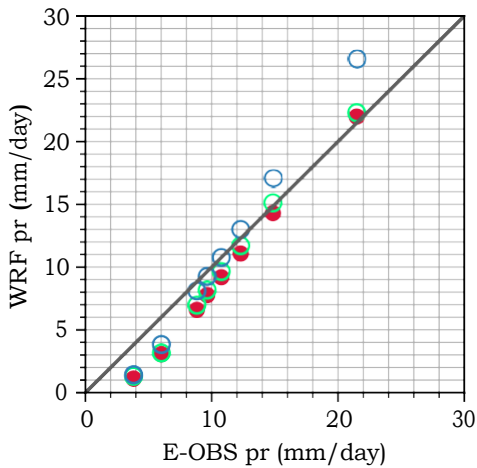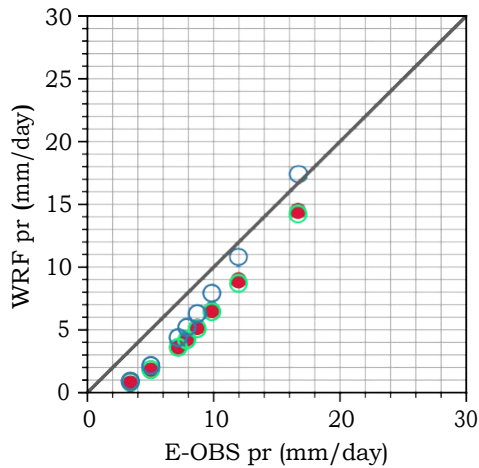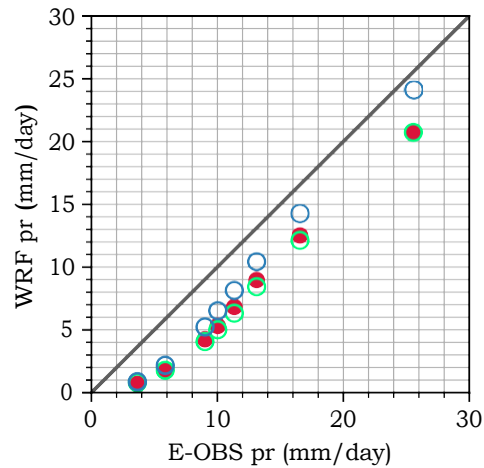

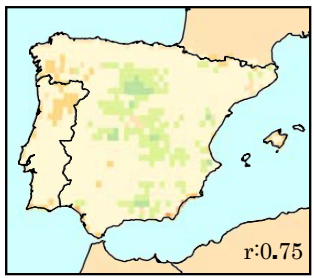 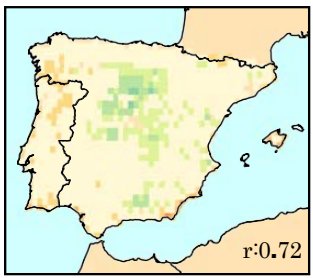 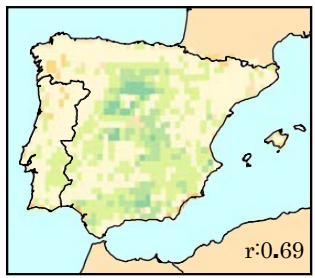
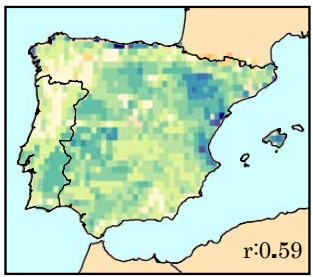 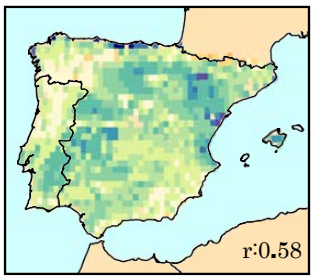 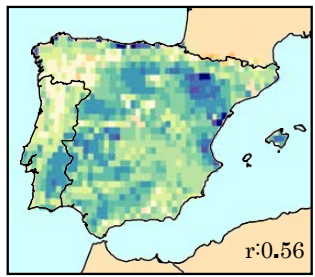
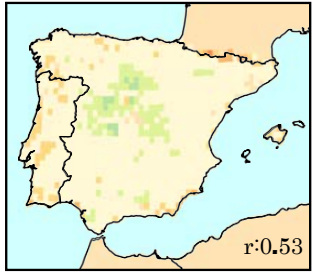 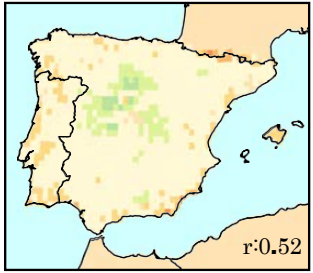 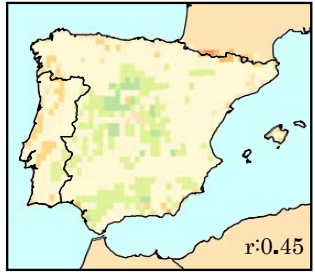
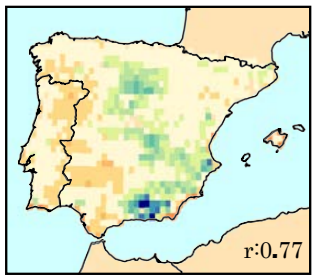 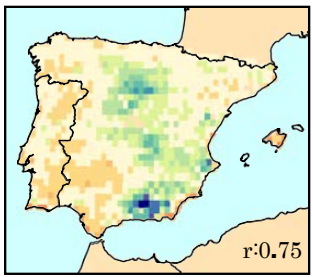 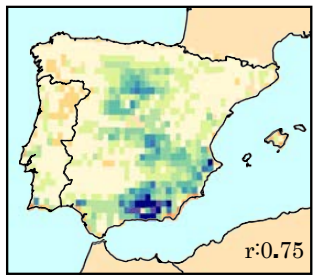
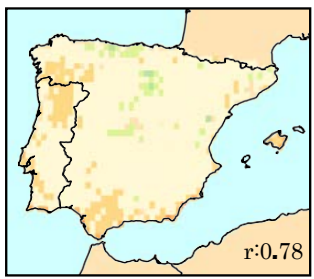 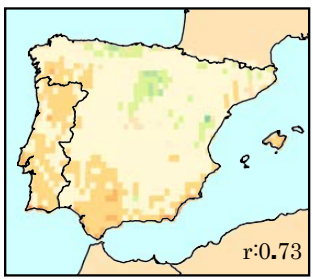 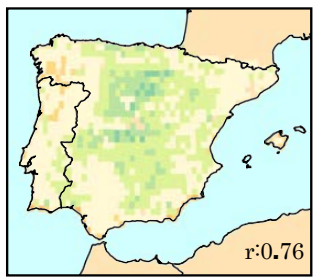
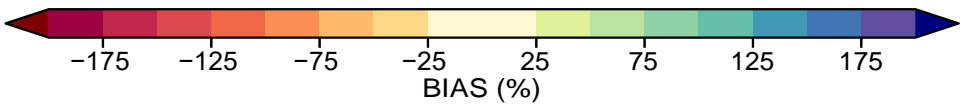

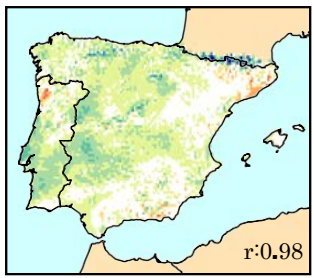
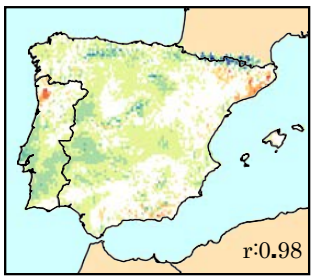
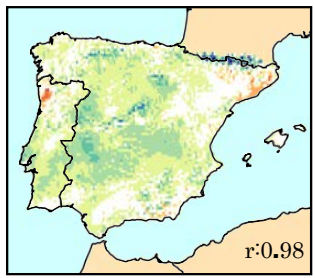
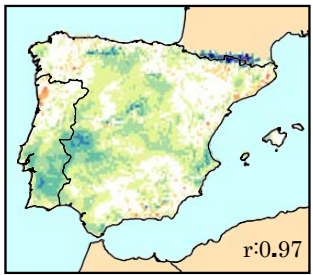
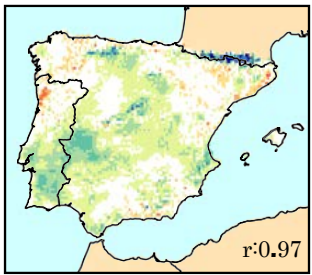
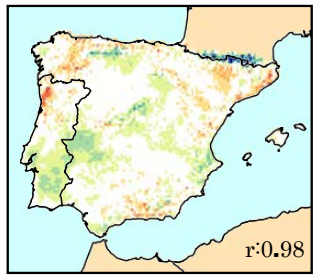
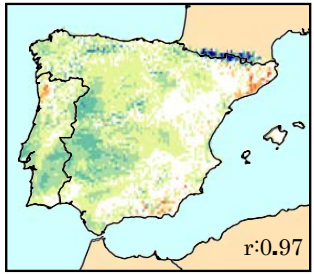
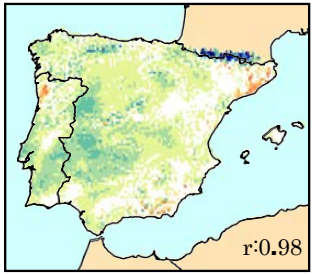
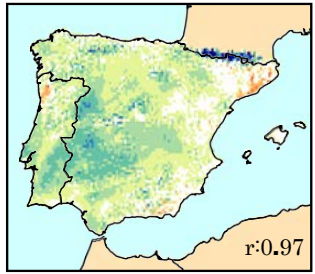
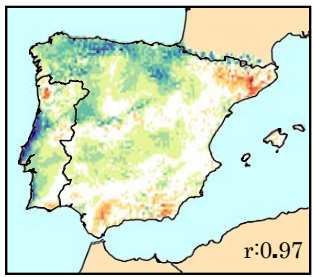
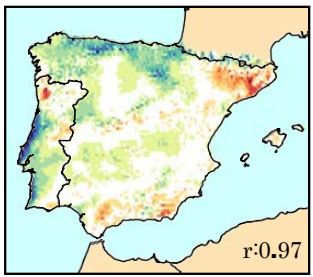
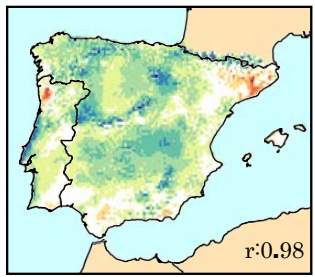
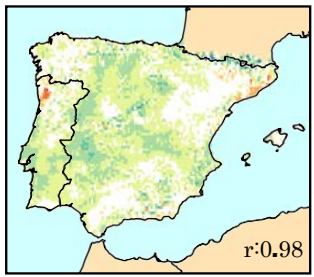
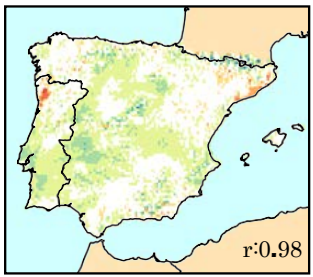
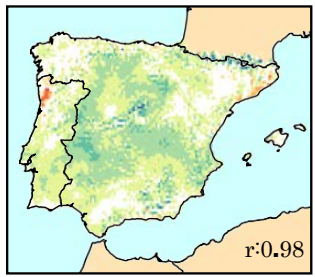
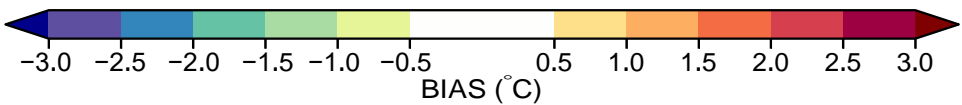

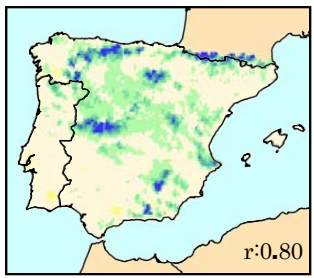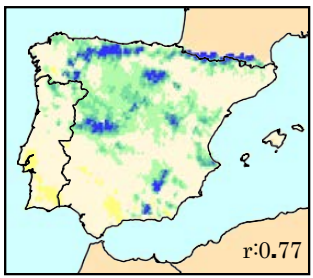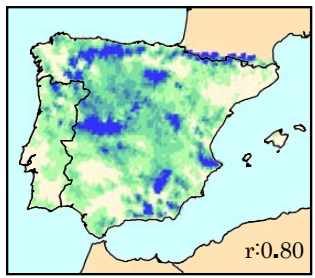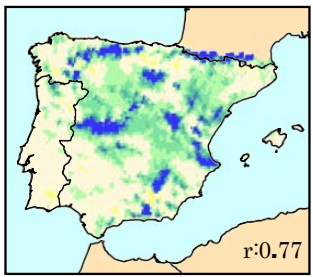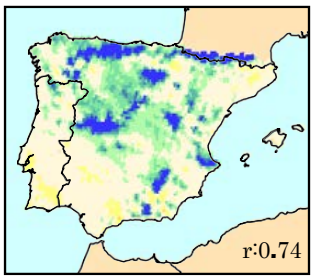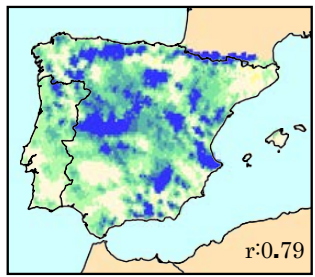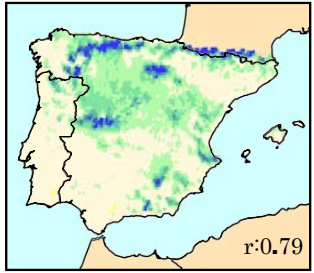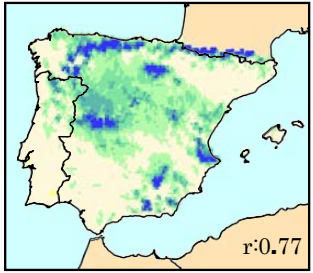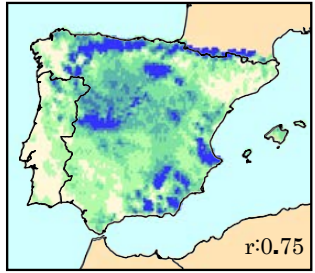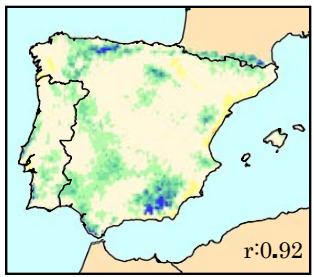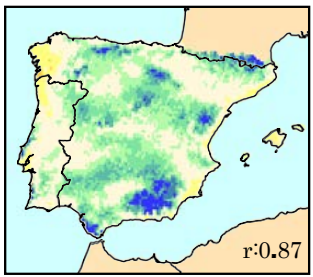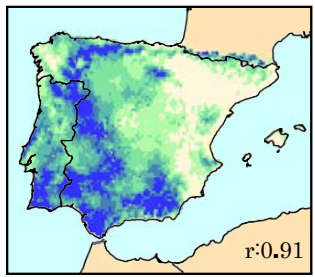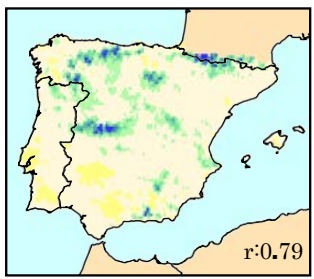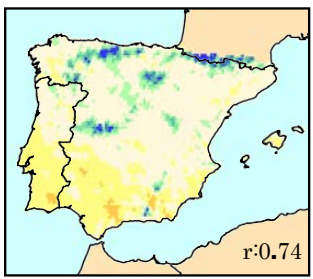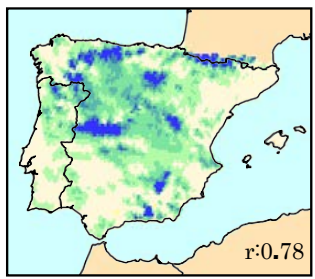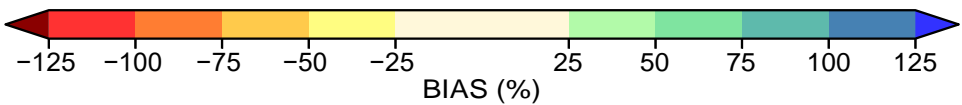

| WRFERA | WRFCCSM | **WRFMPI** |
|---|---|---|

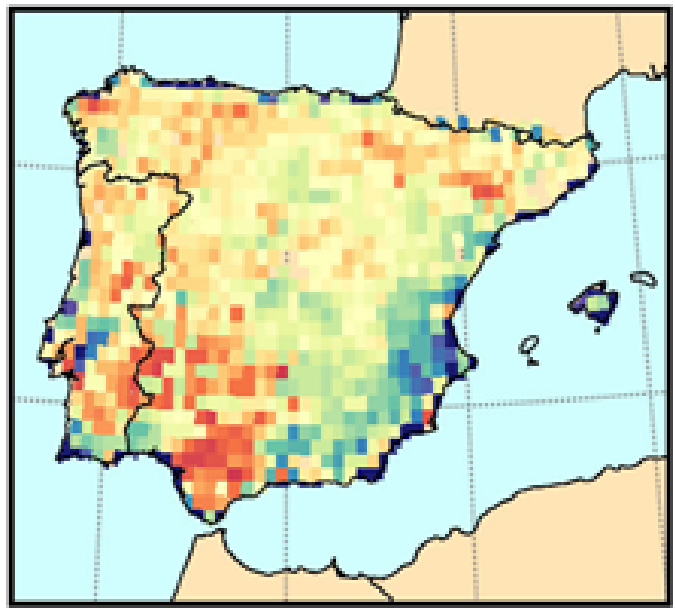 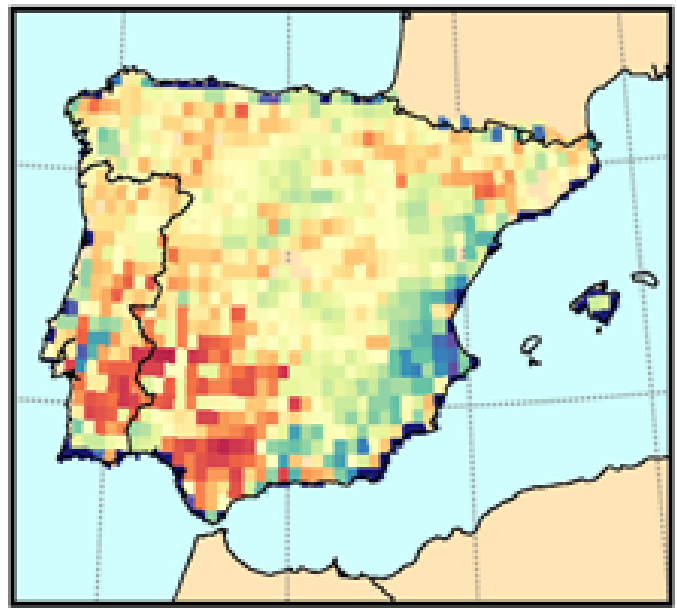 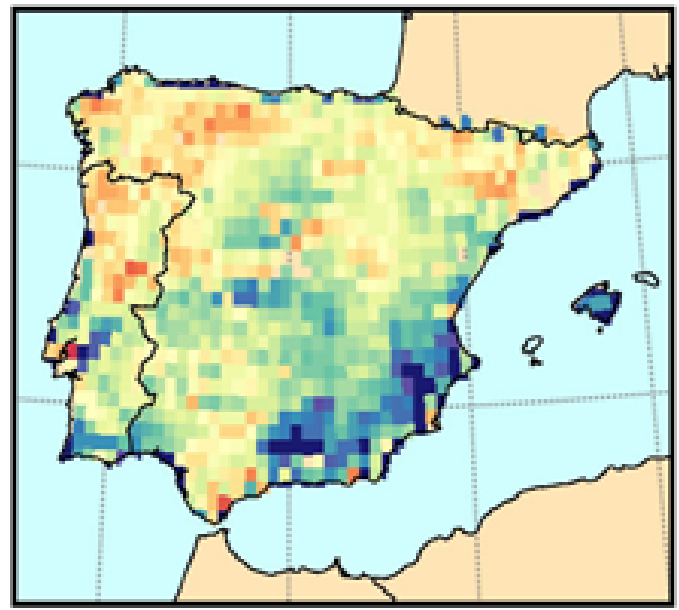

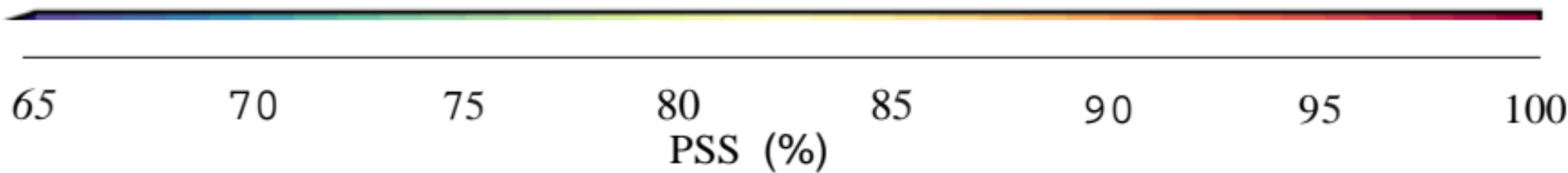

PSS (%)

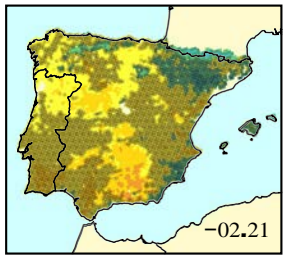 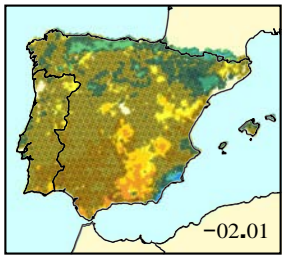 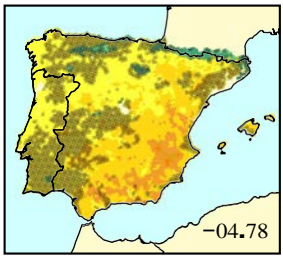 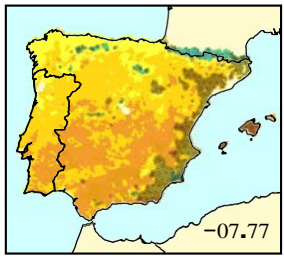
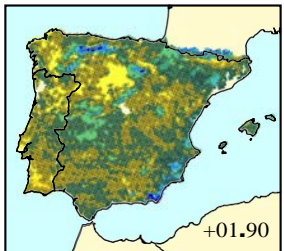 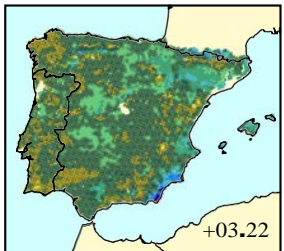 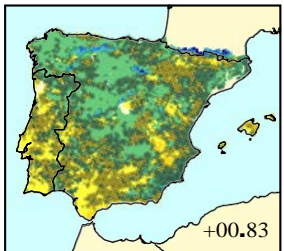 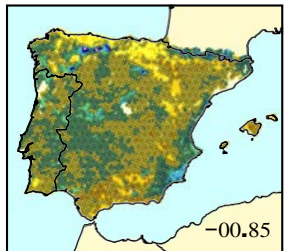
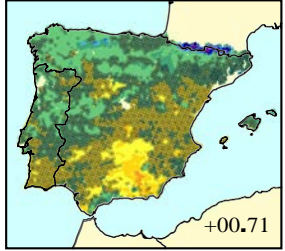 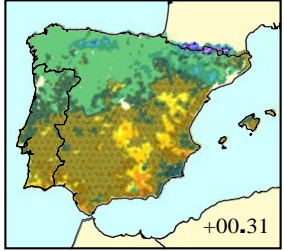 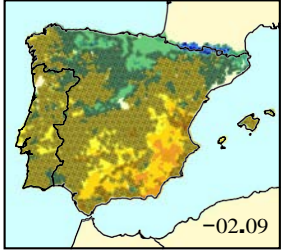 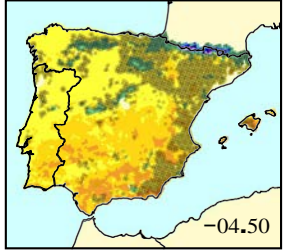
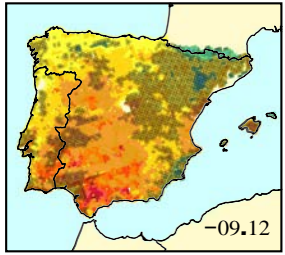 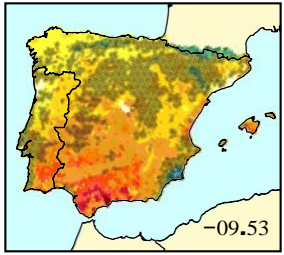 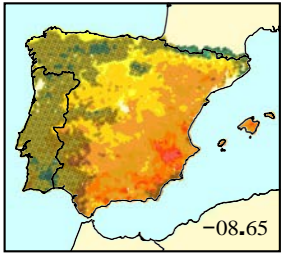 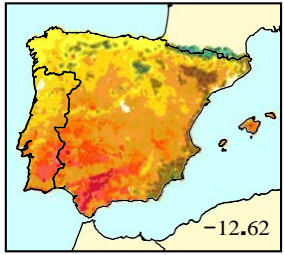
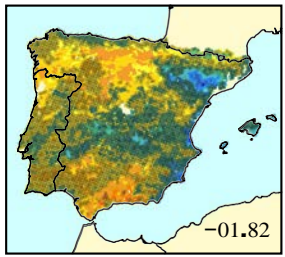 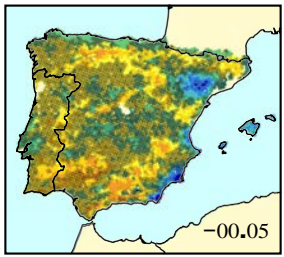 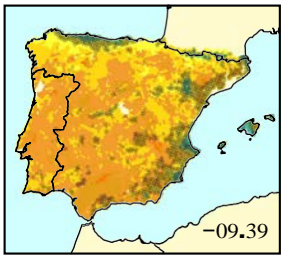 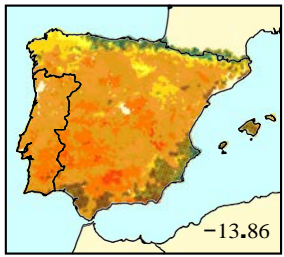
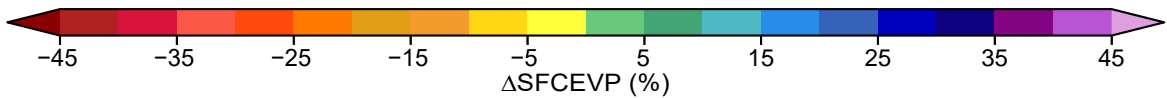

|  | WRFCCSM RCP4.5 | WRFCCSM RCP8.5 | WRFMPI RCP4.5 | WRFMPI RCP8.5 |
|---|---|---|---|---|
| ANNUAL | 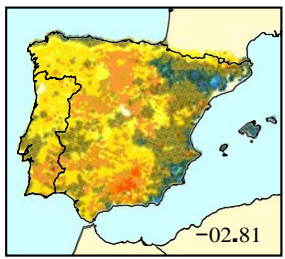 −02.81 | 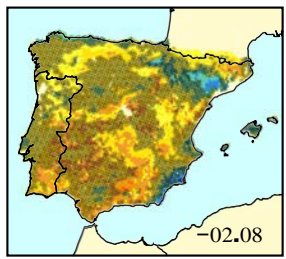 −02.08 | 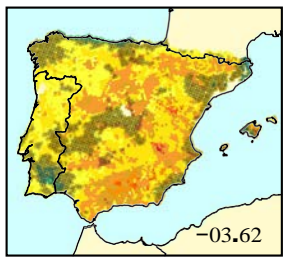 −03.62 | 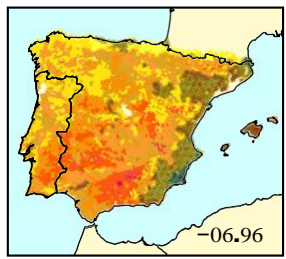 −06.96 |
| DJF | 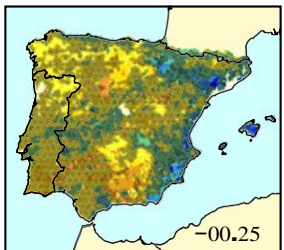 −00.25 | 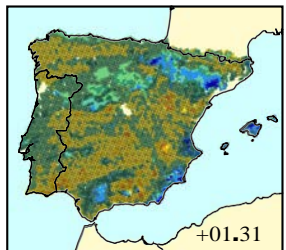 +01.31 | 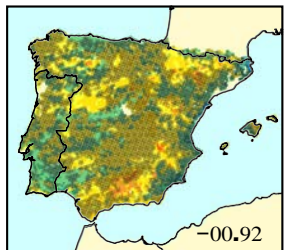 −00.92 | 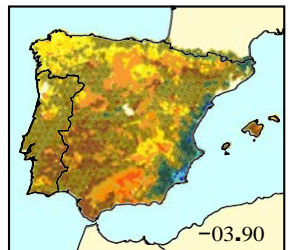 −03.90 |
| MAM | 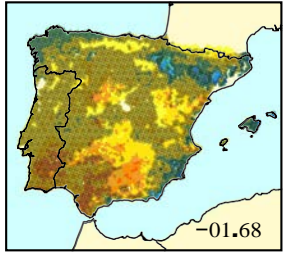 −01.68 | 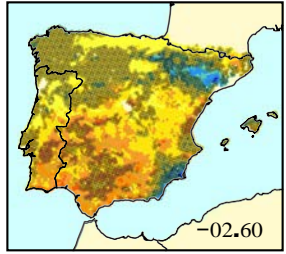 −02.60 | 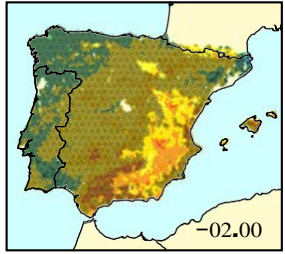 −02.00 | 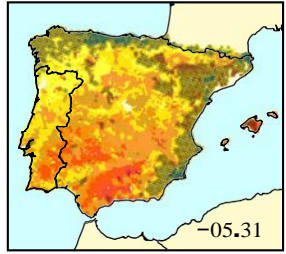 −05.31 |
| JJA | 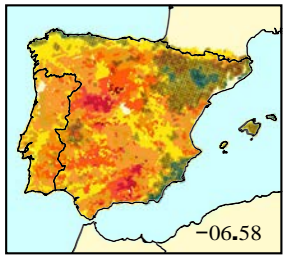 −06.58 | 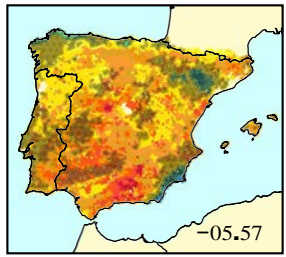 −05.57 | 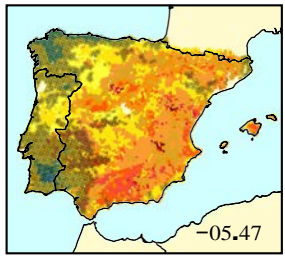 −05.47 | 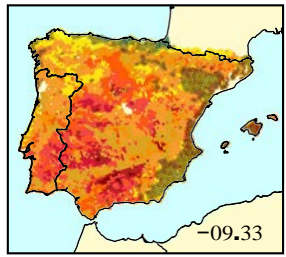 −09.33 |
| SON | 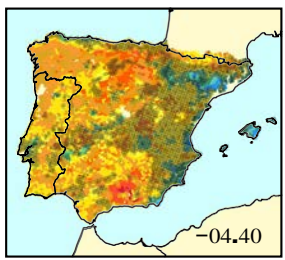 −04.40 | 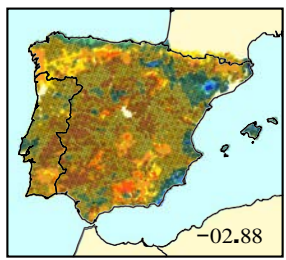 −02.88 | 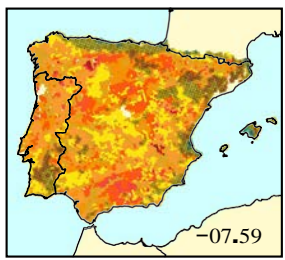 −07.59 | 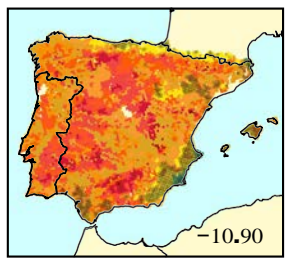 −10.90 |

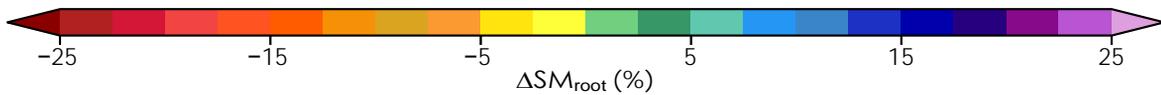

$\Delta SM_{root}$ (%)

**Table 1.** Averaged global temperature rise (ºC) in the near future (2021-2050) period with respect to the historical one (1980-2005), previously linearly detrended (ΔT), and the year at which the temperature rise is above 1.5ºC (+1.5ºC) from the bias-corrected outputs of the two GCMs (CESM1 and MPI-ESM-LR), and under both RCPs (RCP4.5 and RCP8.5).

|  | **CCSM4** | | **MPI-ESM-LR** | |
|---|---|---|---|---|
|  | *RCP4.5* | *RCP8.5* | *RCP4.5* | *RCP8.5* |
| *ΔT (ºC)* | 1.37 | 1.54 | 1.17 | 1.36 |
| *+1.5ºC* | 2039 | 2037 | 2044 | 2039 |

**Table 1.** Monthly error measurements (Bias, MAE, and NormStd) of SFCEVP, T2, and pr for each region (tall vegetation, short vegetation, and urban). Error metrics of simulated (WRFERA, WRFCCSM, and WRFMPI) data were calculated with respect to the reference ones (GLEAM for SFCEVP, and E-OBS for T2 and pr). For SFCEVP and pr, bias, and MAE are depicted in relation to the reference data, and expressed in percentage (%). For T2, bias and MAE are computed in absolute values, and expressed in ºC.

| | **Bias** | | | | | | | | |
|---|---|---|---|---|---|---|---|---|---|
| | *tall vegetation* | | | *short vegetation* | | | *urban* | | |
| | SFCEVP | T2 | pr | SFCEVP | T2 | pr | SFCEVP | T2 | pr |
| *WRFERA* | 12.01 | -0.67 | 33.56 | 1.89 | -0.58 | 18.42 | -96.85 | 1 | 2.33 |
| *WRFCCSM* | 10.91 | -0.53 | 43.96 | 0.58 | -0.46 | 17.94 | -96.87 | 1.13 | 0.19 |
| *WRFMPI* | 17.41 | -0.57 | 69.34 | 16.35 | -0.64 | 54.20 | -96.37 | 1.03 | 28.77 |
| | **MAE** | | | | | | | | |
| | *tall vegetation* | | | *short vegetation* | | | *urban* | | |
| | SFCEVP | T2 | pr | SFCEVP | T2 | pr | SFCEVP | T2 | pr |
| *WRFERA* | 12.61 | 0.69 | 34.98 | 14.42 | 0.60 | 22.19 | 96.85 | 1.00 | 16.27 |
| *WRFCCSM* | 14.40 | 1.34 | 70.66 | 21.37 | 1.29 | 69.48 | 96.87 | 1.47 | 60.51 |
| *WRFMPI* | 19.43 | 1.47 | 88.89 | 25.03 | 1.47 | 90.73 | 96.37 | 1.48 | 73.03 |
| | **NormStd** | | | | | | | | |
| | *tall vegetation* | | | *short vegetation* | | | *urban* | | |
| | SFCEVP | T2 | Pr | SFCEVP | T2 | pr | SFCEVP | T2 | pr |
| *WRFERA* | 0.98 | 0.95 | 1.29 | 0.89 | 1.01 | 1.14 | 0.02 | 1 | 1.07 |
| *WRFCCSM* | 0.96 | 0.96 | 1.33 | 0.88 | 1.01 | 1.05 | 0.02 | 1.01 | 1 |
| *WRFMPI* | 0.98 | 0.92 | 1.54 | 0.93 | 0.96 | 1.44 | 0.03 | 0.97 | 1.28 |

**Matilde García-Valdecasas Ojeda**: Conceptualization, Methodology, Software, Validation, Writing-Original draft preparation.

**Juanjo José Rosa-Cánovas**: Investigation

**Emilio Jiménez-Romero**: Data curation

**Patricio Yeste**: Visualization

**Sonia R. Gámiz-Fortis**: Writing-Reviewing and Editing, Supervision.

**Yolanda Castro-Díez**: Writing-Reviewing and Editing, Supervision.

**María Jesús Esteban-Parra**: Writing-Reviewing and Editing, Supervision, Funding acquisition

# The Role of Surface Evapotranspiration in Regional Climate Modelling: Evaluation and Near-term Future Changes

Matilde García-Valdecasas Ojeda[1], Juan José Rosa-Cánovas[1], Emilio Romero-Jiménez[1], Patricio Yeste[1], Sonia R. Gámiz-Fortis[1], Yolanda Castro-Díez[1] and María Jesús Esteban-Parra[1]

[1]Department of Applied Physics. University of Granada, Granada, Spain

*mgvaldecasas@ugr.es*

# Supplementary figures

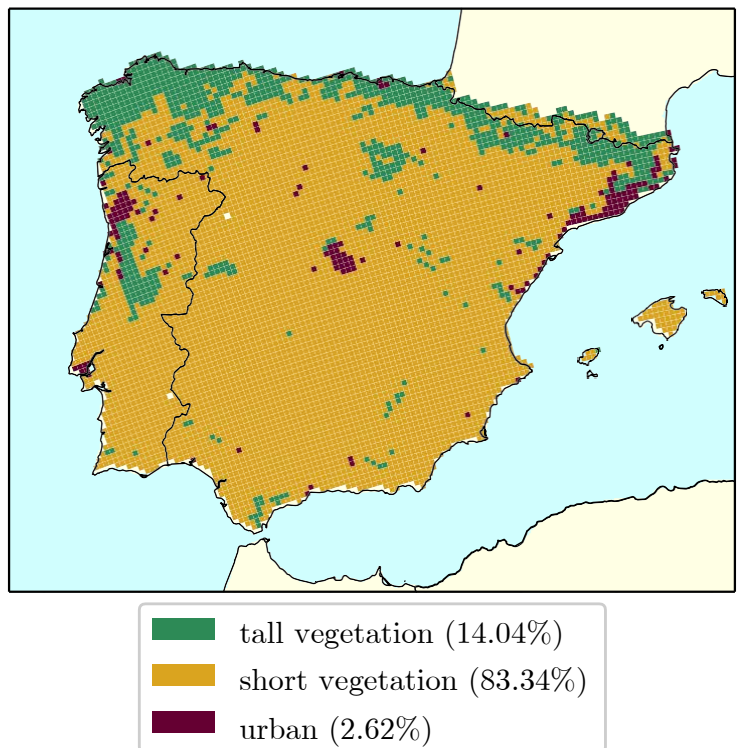

**Fig. 1S** Regions (tall vegetation, short vegetation, and urban) based on the WRF vegetation types. The percentage of coverage is shown in brackets for each region.

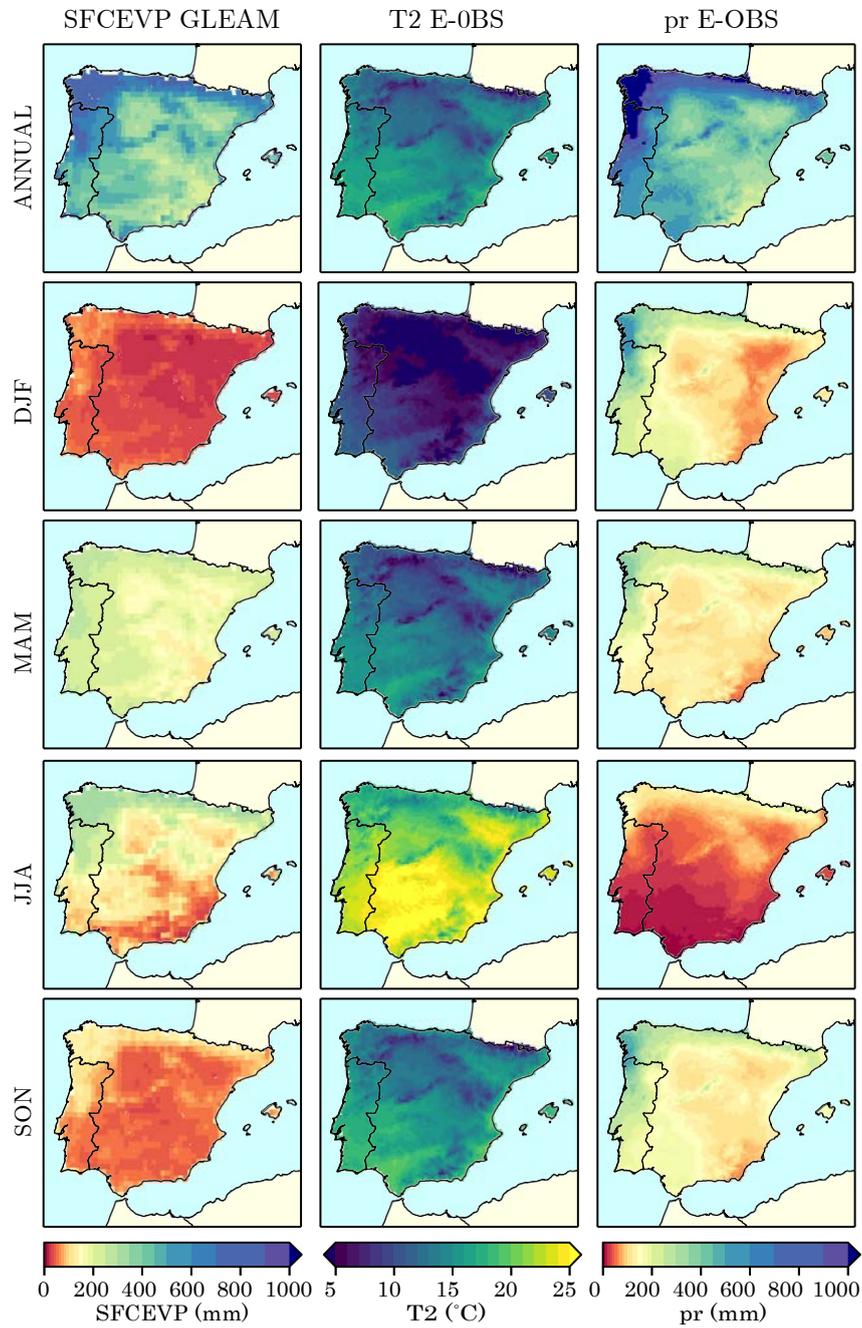

**Fig. 2S** Present-to-day annual (from January to December) and seasonal climatology of the accumulated amount of SFCEVP from GLEAM (first column), averaged T2 from E-OBS (second column), and accumulated pr from E-OBS (third column).

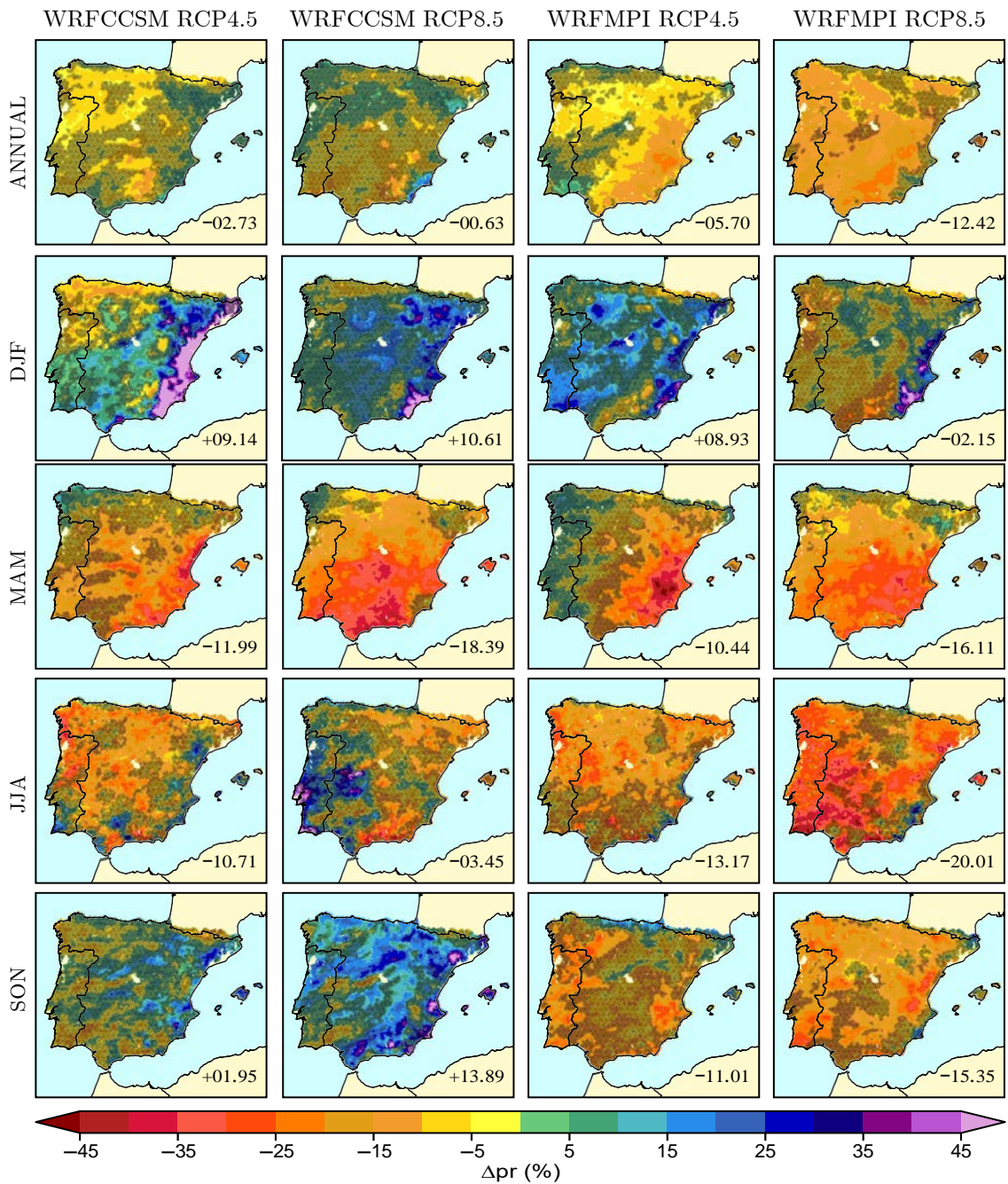

**Fig. 3S** Near-future-to-present relative changes of accumulated pr (%) for the WRFCCSM and WRFMPI simulations under the two RCPs (RCP4.5 and RCP8.5). Stippled areas indicate non-significant changes at the 95% confidence level.

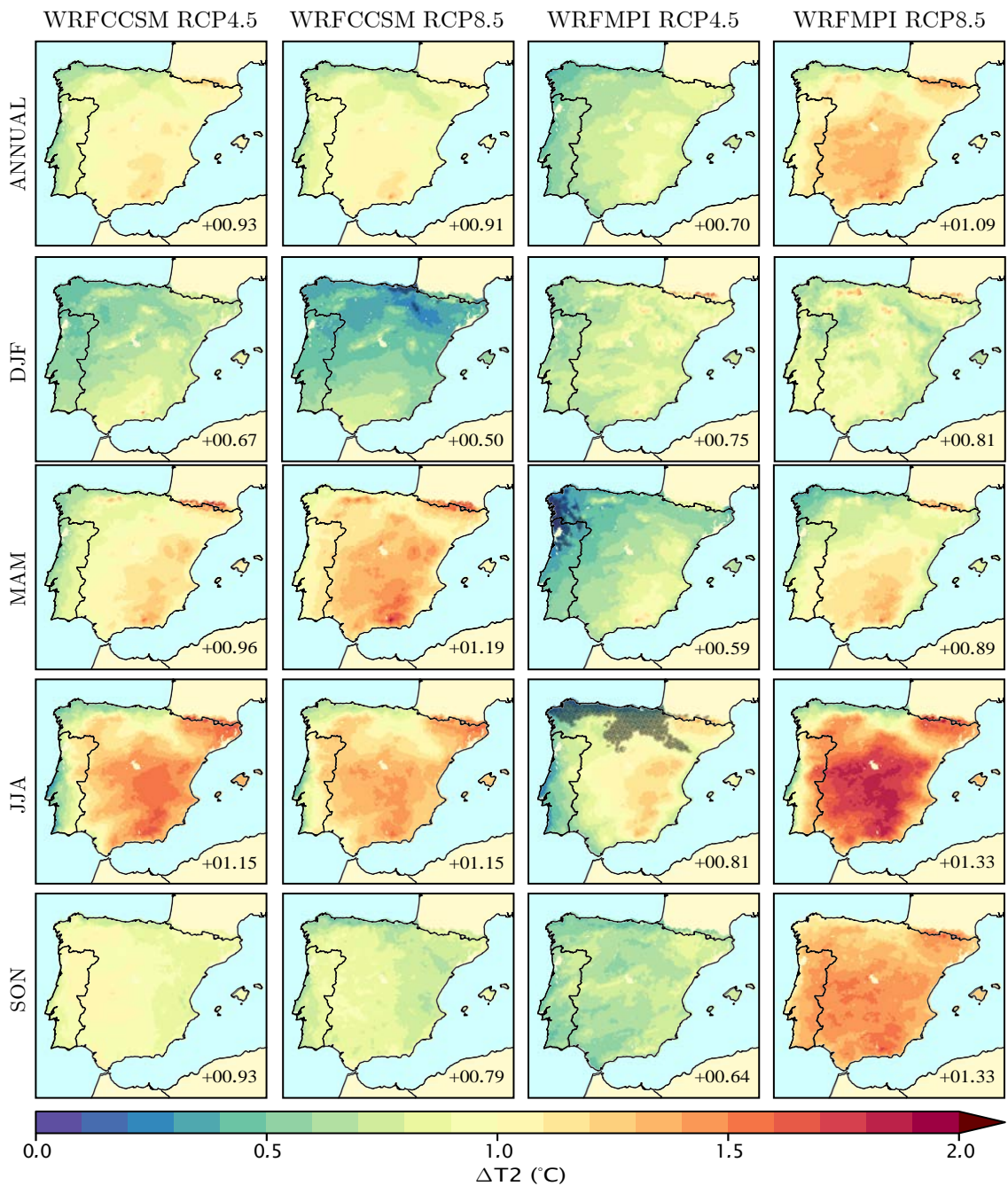

**Fig. 4S** Near-future-to-present changes of average T2 (°C) for the WRFCCSM and WRFMPI simulations under the two RCPs (RCP4.5 and RCP8.5). Stippled areas indicate non-significant changes at the 95% confidence level. The spatial averaged change for the whole IP is indicated in the bottom right corner of each panel.

4